\documentclass[12pt,preprint]{aastex}

\usepackage{amssymb,amsmath}
\usepackage{graphicx}

\begin{document}


\title{Hydrodynamic Simulation of a nano-flare heated \\
    multi-strand solar atmospheric loop}


\author{Aveek Sarkar and Robert W Walsh}
\affil{Centre for Astrophysics,
University of Central Lancashire,  
 Preston, Lancashire, PR1 2HE, UK}
\email{asarkar1@uclan.ac.uk}



\begin{abstract}
There is a growing body of evidence that the plasma loops seen with current instrumentation (SOHO, TRACE and Hinode) may consist of many sub-resolution
elements or strands. Thus, the overall plasma evolution we observe in these features could be the cumulative result of numerous individual strands
undergoing sporadic heating. This paper presents a short ($10^9$ cm $\equiv 10$ Mm) ``global loop'' as 125 individual strands where each strand is modelled
independently by a one-dimensional hydrodynamic simulation. The energy release mechanism across the strands consists of localised, discrete heating
events (nano-flares). The strands are ``coupled'' together through the frequency distribution of the total energy input to the loop which follows a power
law distribution with index $\alpha$. The location and lifetime of each energy event occurring is random. Although a typical strand can go through a
series of well-defined heating/cooling cycles, when the strands are combined, the overall quasi-static emission measure weighted thermal profile for the global loop reproduces a
hot apex/cool base structure. Localised cool plasma blobs are seen to travel along individual strands which could cause the loop to `disappear' from coronal emission and appear in transition or chromospheric ones. As $\alpha$ increases (from 0 to 2.29 to 3.29), more weight is given to the smallest heating episodes. Consequently, the
overall global loop apex temperature increases while the variation of the temperature around that value decreases. Any further increase in $\alpha$
saturates the loop apex temperature variations at the current simulation resolution. The effect of increasing the number of strands and the loop length
as well as the implications of these results upon possible future observing campaigns for TRACE and Hinode are discussed.

\end{abstract}


\keywords{Sun - activity; Sun - corona; hydrodynamics - Sun}



\section{Introduction}
The discovery that a significant proportion of the radiation emitted from the solar corona is concentrated along well-defined loops represented a major advance in our understanding of the Sun. These loops are the basic structural elements of the atmosphere with the Solar and Heliospheric Observatory (SOHO), Transition Region and Coronal Explorer (TRACE) and the Hinode missions revealing them in unprecedented detail. It is now believed universally that these features coincide with magnetic flux tubes and occur as the plasma and thermal energy can flow along but not easily across the magnetic field.
However it must be noted that the phrase ``loop'' is an inclusive, general term. In particular, there is the discussion as to whether an individual loop has yet to be resolved; that is, do even TRACE EUV loops further consist of a bundle of filamentary plasma strands at a range of temperatures which, when averaged over, give the appearance of  uniformly bright structures (Lenz  et al., 1999)? Using SOHO-CDS \& Yohkoh soft X-ray observations for 13 positions along a given loop structure, Schmelz et al. (2001) argue that the resulting broad Differential Emission Measure (DEM) is a strong indicator of a multi-thermal plasma. Since the heat transport across the magnetic field should be very small, the conclusion could be that the loop under investigation is multithreaded in nature. 

One dimensional (1D) hydrodynamic (HD) modelling of a loop as plasma evolving along individual field-lines has been popular since the late 1970's (Peres, 2000; Walsh et al., 1995). However, it must be recognised that if the range of loop structures we can observe do consist of many ``sub-resolution'' elements, then these 1D models are really only applicable to an individual plasma element or strand. Thus, a {\bf{loop}} is an amalgamation of these {\bf{strands}}. They could operate in thermal isolation from one another with a wide range of temperatures occurring across the structural elements.

Up until now, several multi-strand static models have been associated with specific observations. For example, Reale \& Peres (2000) showed that their multi-strand (6) hydro-static solution is in rough agreement with an isothermal loop observed by TRACE. Aschwanden et al. (2000) compares hydrostatic solutions with 41 TRACE EUV loops of different lengths; mostly they fall short of being able to reproduce the TRACE emission (the loops appear to be denser that those generated by static calculations). To explain the discrepancy, Winebarger et al. (2003) concludes that it is unlikely that the over-dense loops can be reconciled with any static model.

 Warren et al. (2002) outline a multi-strand loop model by choosing ten randomly selected time periods from their single hydrodynamic loop simulation, synthesise TRACE 171 and 195 $\mathring{A}$ intensities and average the resulting emission over the threads. They obtain  a flat 171/195 filter ratio along the loop resulting in much larger coronal intensities relative to those estimated by static heating. Ugarte-Urra et al. (2006) employ an impulsive and quasi-static heat input to a hydrodynamic loop model to examine the subsequent evolution in X-ray and EUV; they find that compared to observed loops, the simulated EUV response lifetime for a single cooling loop is much shorter than those observed from TRACE. Also, Warren (2006) employs a series of multi-strand (50) loop simulations to reproduce the high temperature evolution of solar flare: the numerical results suggest that an individual strand  has an optimum heating timescale of a few hundred seconds.

Thus this leads onto another important question of how the million degree plasma within loops is heated in the first place. One of several possible theoretical heating mechanisms is the concept that the plasma is energised by the cumulative effect of numerous, small scale ($\sim 10^{24}$ erg per event), localized, time-dependent energy bursts or nano-flares (Parker, 1988). It is observed already that for larger solar flares their frequency of occurrence $f$ has a dependence upon the energy content ($E$) and that it follows a power law;
\begin{equation}
df/dE=E_0E^{-\alpha},
\end{equation}
with an index of $\alpha \sim 1.8$. Hudson (1991) pointed out that for the corona to be heated predominantly by nano-flares, a steeper-slope ($\alpha > 2$) would be required. Several authors claim to have observed this steeper distribution from observed brightenings in both EUV and X-rays (eg. Pauluhn \& Solanki 2007, Krucker \& Benz 1998, Parnell \& Jupp 2000).

Thus, if nano-flare heating is taking place within loops, then multiple sub-resolution strand modeling with a heat input that is episodic in nature should be important. One approach to simulating this scenario is to use a zero-dimensional (0D) hydrodynamic calculation as introduced by Cargill (1994) and later modified by Cargill \& Klimchuk (1997, 2004) and Klimchuk \& Cargill (2001). These authors devised a semi-analytic, multi-strand model where it is assumed that each strand can be represented by a single temperature and density only. Each strand experiences ``impulsive'' nano-flare heating in the sense that the heat deposition occurs on timescales much shorter than any plasma cooling time. The heated plasma cools initially by conduction and then later by radiation. Subsequently, a ``global loop'' is constructed of many (500-5000 say) strands and observables (eg. emission measure) were calculated. The results show that increasing the number of strands in the global loop leads to a slight increase in overall average temperature but the emission measure remains almost unaffected. The model also explains the overdensity of the warm coronal loops (Cargill \& Klimchuk, 2004).

Following on from this model, Cargill \& Klimchuk (1997) compared the radiative signature of their 0D loop with Yohkoh Soft X-ray telescope (SXT) observations. Observed loop dimensions and radiative losses were used as multi-strand nano-flare model inputs and observables such as temperature, emission measure and filling factors were derived. Their results show that the model agrees fairly well with very hot loops ($T> 4 \times 10^6$ K) but not for cooler loops ($<T \sim 2 \times 10^6$ K). Subsequently in Cargill \& Klimchuk (2004), the authors study the effect of altering the power law index in their nano-flare energy distributions. Their results show that steeper power law indices (eg. $\alpha = 4$) changed considerably the emission measure profiles and the value of the filling factor, compared to the flat ($\alpha=0$) distribution.

In that regard, Patsourakos and Klimchuk (2005) generate synthetic line intensities from a  nano-flare heated hydrodynamic loop simulation. They localise the spatial distribution of the nano-flare events, finding that the resulting TRACE and Yohkoh SXT emission was only affected weakly by the various dominant heat deposition locations. Also, Patsourakos and Klimchuk (2006) stress the importance of predicting line profiles for their nano-flare-heated loop model, indicating that the profile for a hot line (in this case Fe XVII at $\approx 5$ MK) should be seen to undergo strong broadening  with distinctive enhancements  in the line wings. 

This current paper extends greatly the above by examining a fully 1D hydrodynamic simulation of a small ($10$ Mm) loop which consists of 125 individual strand elements. Each strand operates independently in regard to the plasma response along the structure; however, the strands are connected through the frequency distribution of the energy input via small scale heating episodes which follows a power law for a predefined index. The paper is arranged as follows: in Section 2, the numerical model for a single strand is outlined as well as the plasma response to the sudden deposition of a nano-flare sized energy burst. Section 3 constructs a `global loop' consisting of multiple (125) strands where each strand is subjected to several successive impulsive energy bursts. Subsequently, the effect of varying the power law index $\alpha$ is investigated. Finally, Section 4 presents the Discussion and an outline of future work (both through further simulation and possible observations).  \\

\section{Single strand model}
Consider a $10^9$cm $=10$ Mm long loop with a cross-sectional radius of $\backsim 1.1 \times 10^8$ cm $=1.1$ Mm. Let us assume that this loop consists of 125 individual plasma strands which fill the loop volume (that is, the radius of each strand is $9.6 \times 10^6$ cm $=0.098$ Mm). These strands are thermally independent so that the dynamics of one strand can not affect any other. The evolution of an individual strand in response to a designated heat input are outlined in the following.

\subsection{Numerical model of a strand}

Given the fact that the solar corona is a highly conducting low-$\beta$ medium, the magnetic field confines the plasma along flux tubes, and the plasma can be described with one-dimensional hydrodynamics. A Lagrange-remap one-dimensional hydrodynamic code (adapted from Arber et al. (2001)) is employed for the purpose of solving the following time dependent one dimensional differential equations of mass, momentum and energy conservation; 

\begin{equation}
\dfrac{D \rho}{D t}+\rho\dfrac{\partial}{\partial s}v=0
\end{equation}
\begin{equation}
\rho \dfrac{Dv}{Dt}=-\dfrac{\partial p}{\partial s}+\rho g+ \rho \nu \dfrac{\partial ^ 2 v}{\partial s^2}
\end{equation}
\begin{equation}
\dfrac{\rho^\gamma}{\gamma -1}\dfrac{D}{Dt}(\dfrac{p}{\rho^\gamma})=\dfrac{\partial}{\partial s}(\kappa\dfrac{\partial T}{\partial s})-n^2 Q(T) + H(s,t)
\end{equation}

\begin{equation}
p=\dfrac{R}{\tilde{\mu}}\rho T
\end{equation}

\begin{equation*}
\dfrac{D}{Dt}\equiv \dfrac{\partial}{\partial t}+ v\cdot \dfrac{\partial}{\partial s}
\end{equation*}

where $\rho, p, n, \textbf{v}$ and $T$ represent mass density, pressure, particle density, velocity and temperature of the plasma respectively. $s$ is the spatial co-ordinate along the strand ($-L<s<L$ where in this case $L=5$Mm) which is assumed to be semi-circular. In Equation (3), $g$ represents the component of the gravity along the semicircular loop; because we are considering a small loop, we assume $g$ to be a constant of value equal to the surface value ($2.74 \times 10^4$ cm s$^{-2}$) for all points along the loop. $\gamma$ represents the adiabatic index of the medium, which we consider to be $5/3$. $\kappa$ is the conductivity of the plasma in the direction of $s$ ($= 9.2 \times 10^{-7}T^{5/2}$ erg s$^{-1}$ cm$^{-1}$ K$^{-1}$). $R$ the molecular gas constant ($8.3 \times 10^7$ erg mol$^{-1}$K$^{-1}$) and $\tilde{\mu}$ is the mean molecular weight with $\tilde{\mu}=0.6$ mol$^{-1}$. The coefficient of kinematic viscosity $\nu$ is assumed to be uniform throughout the plasma. $Q(T)$ is the optically thin radiative loss function; we adopt a piecewise continuous function having general form $Q(T)=\chi T^{\beta}$, which is based on the work by Rosner, Tucker and Vaiana (1978). $H(s,t)$ is our prescribed spatially and temporally dependent coronal heating term. The coronal loop is assumed to be symmetrical and initially, 
\begin{equation}
\dfrac{\partial T}{\partial s}=\dfrac{\partial p}{\partial s}=0,
\end{equation}

at the loop apex ($s=0$). The boundary conditions follow as;
\begin{equation}
T(-L,t)=T(L,t)=T_{ch}=10^4 \; \text{K}
\end{equation}
\begin{equation}
p(-L,t)=p(L,t)=p_{ch}=0.314 \; Pa
\end{equation}
at the loop footpoints deep in chromosphere (where $T_{ch}, p_{ch}$ are the chromospheric temperature and pressure respectively). The chromosphere has a depth of $0.4$ Mm at each loop leg. At the beginning of the simulation the temperature along the strand is kept at the chromospheric temperature i.e. $10^4$ K and the velocity along the strand is kept fixed at zero. The pressure as well as the density is decreased exponentially towards the strand apex; Subsequently; the plasma is gravitationally stratified and higher density plasma at the chromosphere is available for chromospheric evaporation during the simulation.

 As we shall see in these simulations, the sudden release of the localised energy bursts will create travelling shock fronts throughout the strand plasma. The Lagrange remap code has been shown to deal very well with resolving these type of fronts (Arber et al. 2001). Thus, an average grid spacing of $0.037$ Mm was employed at the central coronal part of the loop; this optimises the simulation in terms of both the resolution required to track the dynamic features in the strand versus a reasonable simulation run time. \\
 
\subsection{Strand plasma response to a nano-flare} 
\begin{figure}
\centering
\includegraphics[width=0.45\textwidth,angle=0]{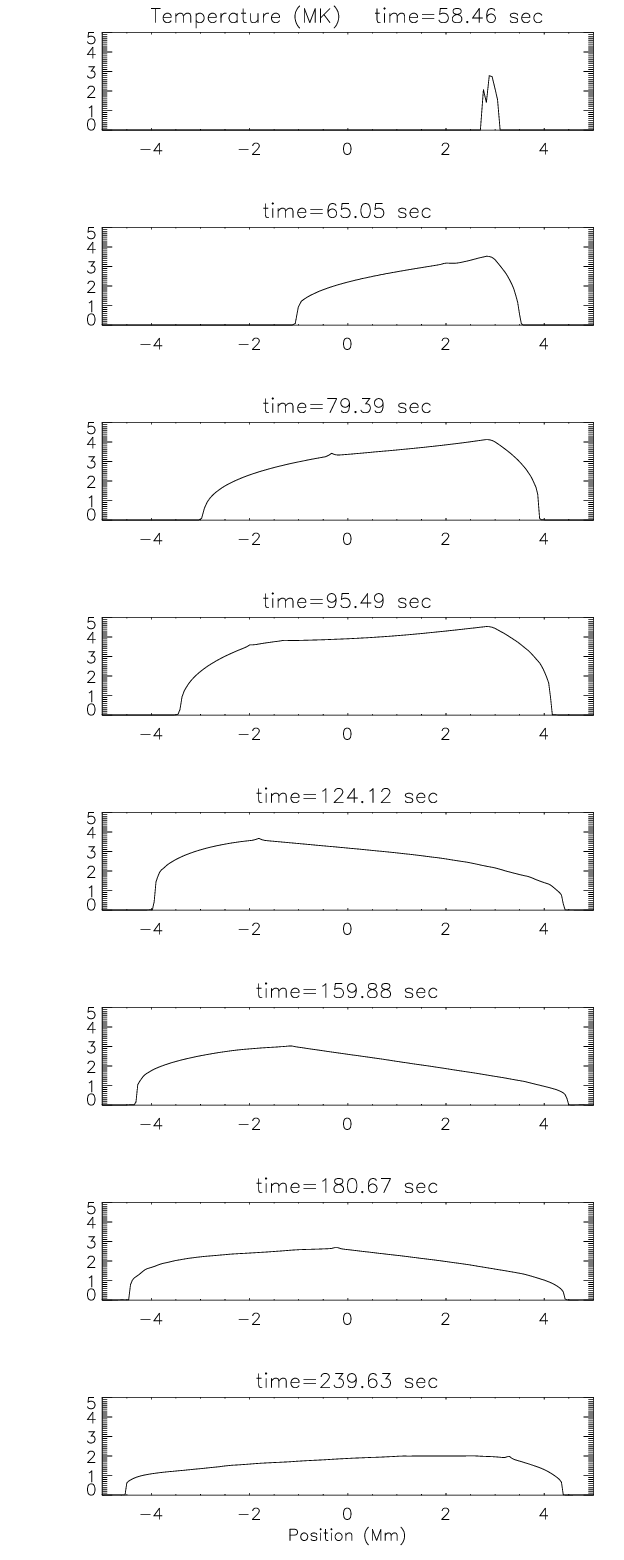}
\includegraphics[width=0.45\textwidth,angle=0]{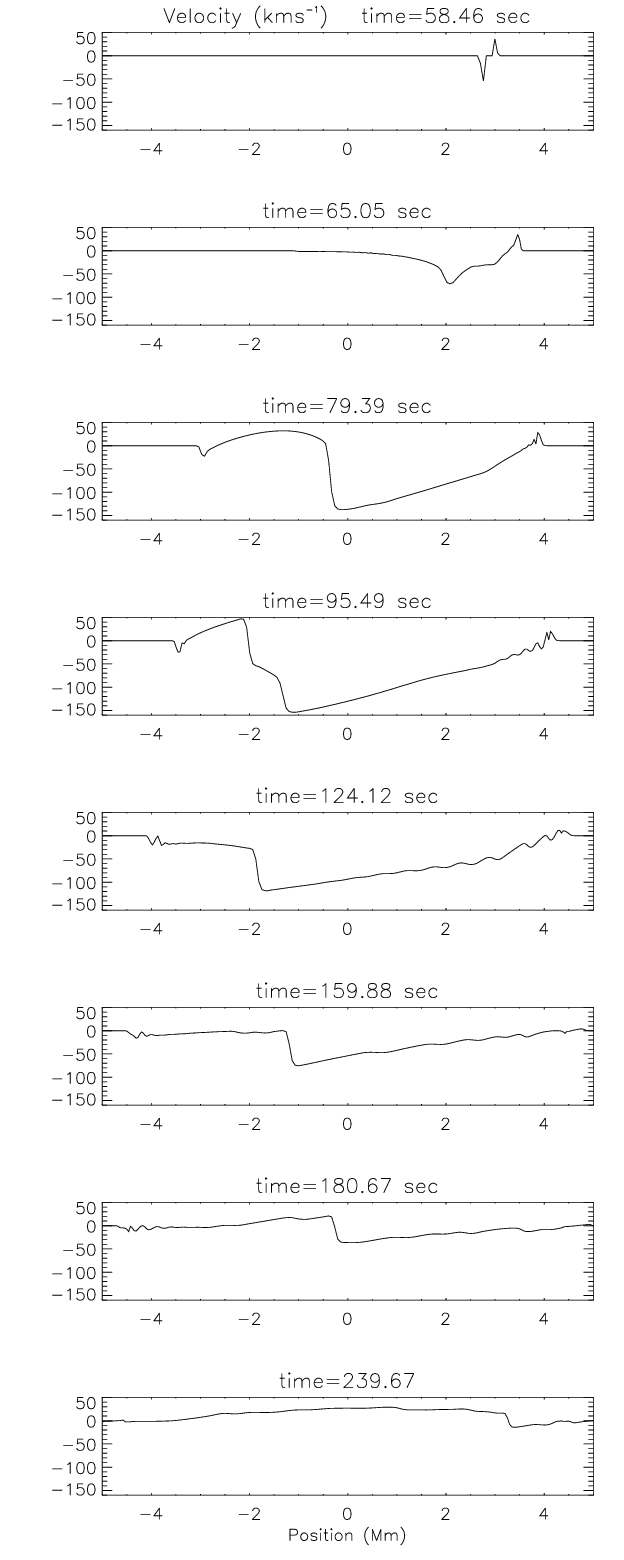}
\caption{Successive snapshots of the strand evolution in temperature (left) and velocity (right) in response to $1.049 \times 10^{24}$ erg of energy deposited at $2.8$ Mm to the right of the strand apex (at $s=0$). \label{plotone}}
\end{figure}

Using the above strand model, let us examine the response of the plasma within the initially cool, evacuated strand to a rapid deposition of localised energy - a nano-flare. Consider the evolution of the plasma temperature and velocity along the strand as shown in Figure (\ref{plotone}). Here, the total energy contained within the heating burst is $1.049 \times 10^{24}$ erg. The heating is localised at $2.8$ Mm to the right of the strand apex (at $s=0$) and occurs over a lengthscale of $0.2$ Mm. The event lifetime is $50$ seconds (starting at 57.5 seconds after the simulation begins). \\

After an initial localised raising of the temperature where the energy is deposited, the extra heat is carried away from this site by conduction; eventually the overall strand temperature rises up to over $4$ MK. From the velocity snapshots it is clear that, due to sudden heating, a shock front develops (snapshot $58.46$ sec and $65.05$ sec) that propagates along the strand with a velocity up to over $140$ km $s^{-1}$. As expected from basic acoustic shock front physics, the propagation of the front is also observed in a slight local increase in the temperature.  At time $107.5$ second, the heating burst is switched off. Sound waves continue to bounce back and forth along the strand (reflecting off the high density chromospheric boundary) and eventually the overall temperature begins to decrease, the shock front decays and the plasma velocity declines (to $30$ km $s^{-1}$ at $239.63$ s).

\begin{figure}
\centering
\includegraphics[width=0.35\textwidth,angle=90]{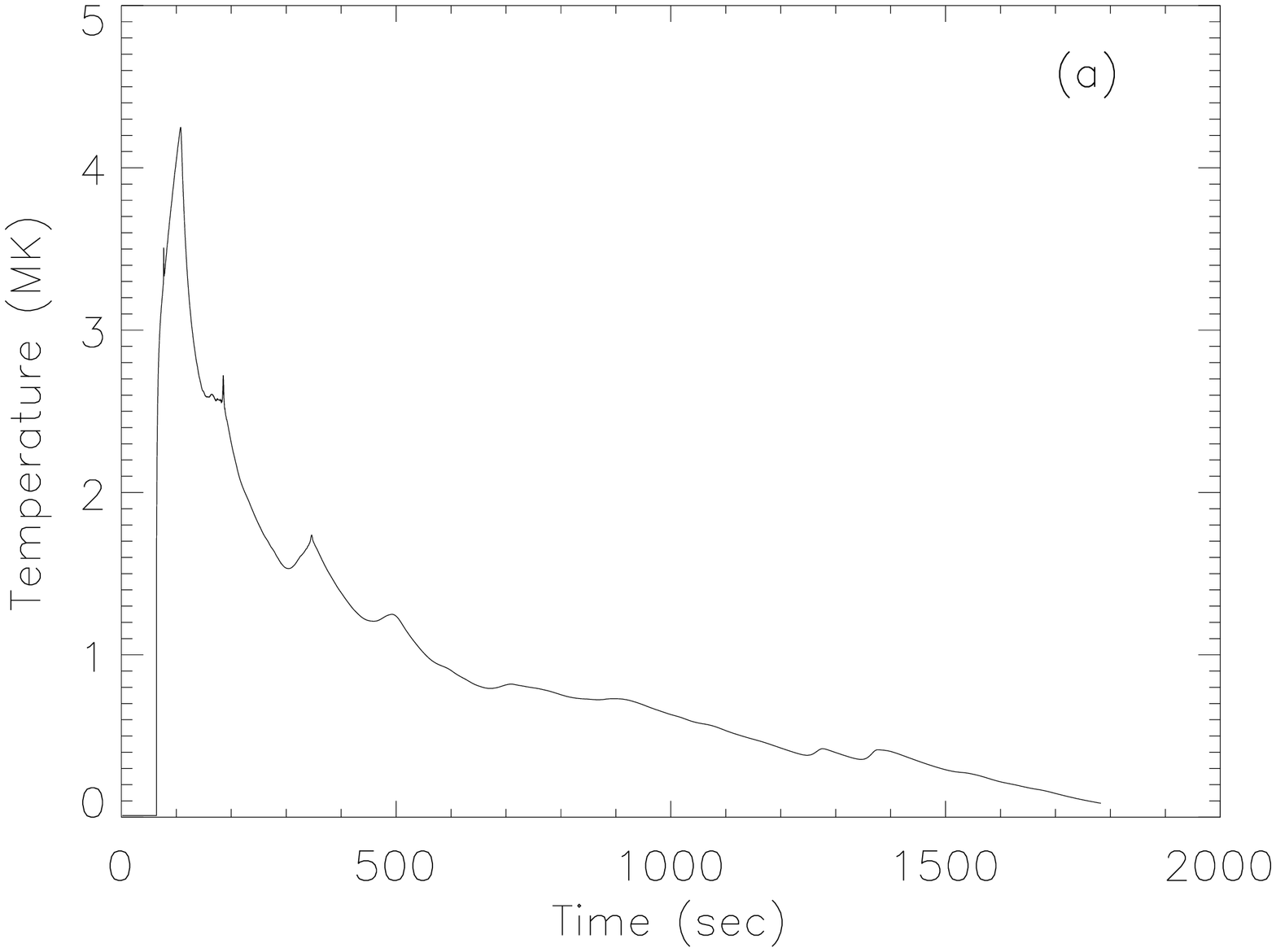}
\includegraphics[width=0.35\textwidth,angle=90]{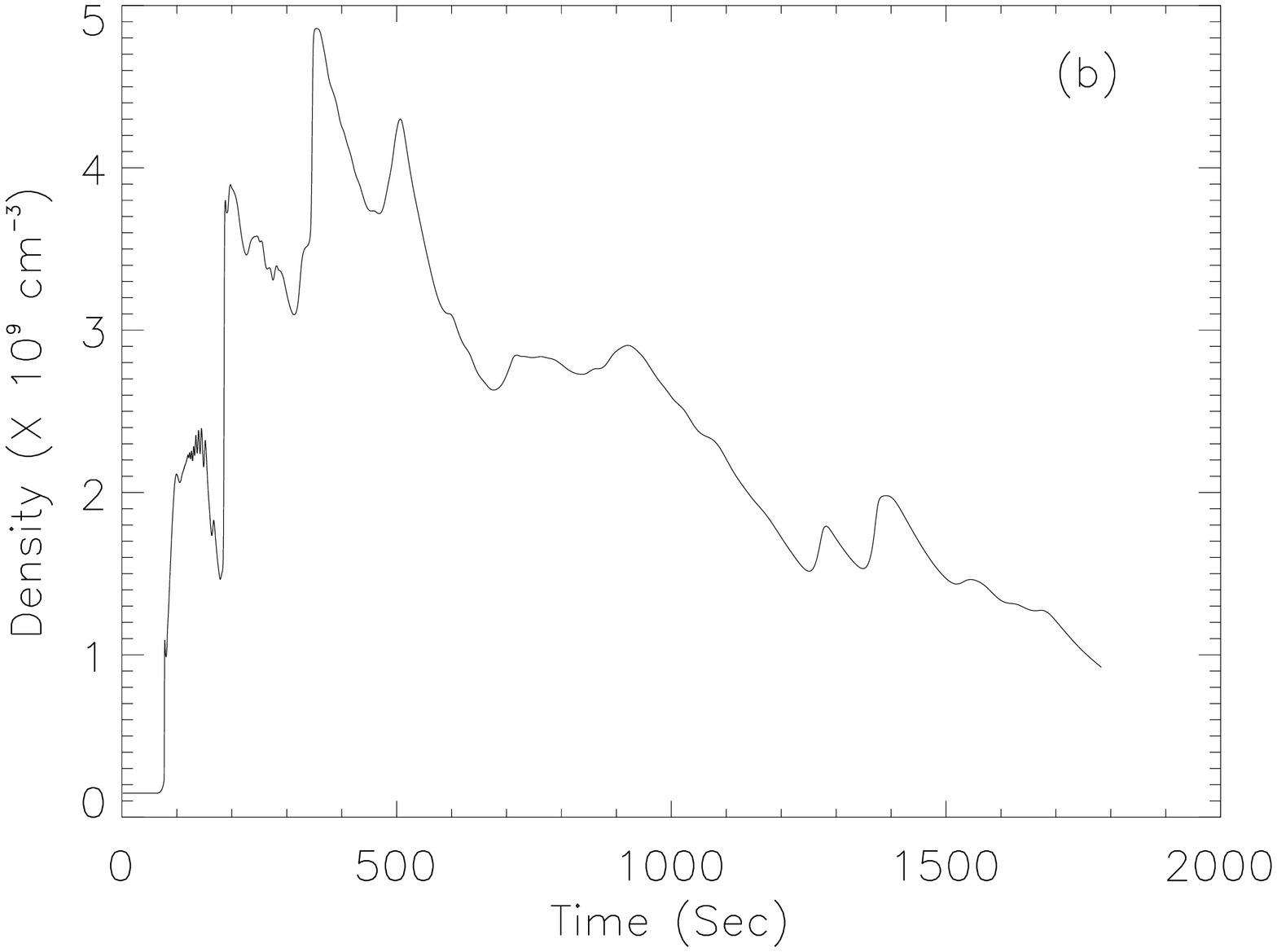}
\caption{Evolution of strand-apex (a) temperature and (b) density, after a localised energy burst.  \label{plottwo}}
\end{figure}

   The main feature of previously described 0D model is that initially, it allows a strand to cool solely by conduction until the ratio of conduction time scale to radiation time scale becomes unity. Thereafter, radiative losses take over. In contrast, the present model keeps both processes active with the dominate cooling mechanism determined automatically when solving the set of hydrodynamic equations. In addition to that the present model is capable of transporting localised extra heat by means of mass flow via enthalpy flux. Thus consider Figure \ref{plottwo} where the strand apex (a) temperature and (b) density are seen to evolve after the nano-flare burst. The density evolution shows clearly that chromospheric evaporation continues to take place up to around $300$ s, after that density drops as the plasma condenses back to the chromosphere. In contrast to Figures 1 \& 2 in Cargill (1994), there is no sudden switch between cooling timescales. Also, there is some smaller-scale structures due to the flow of material along the loop as the plasma cools; the spatial evolution of this plasma flow is shown already in the Figure \ref{plotone}.

\section{Multi-strand model}
Consider now a loop as consisting of 125 individual strands where these strands are related through the distribution of the localised energy input across the elements. The localised heat input $H(s,t)$ arising from the deposition of a given amount of energy $E$ over an event lifetime $\tau$ are chosen randomly in time with constraints $10^{23}erg \leq E \leq 5 \times 10^{24}erg$ and $50s \leq \tau \leq 150 s$. 

The energy bursts are released within a fixed volume element where the element length is $0.2$ Mm. The heating episode location ($S_L$) can take place anywhere along the strand within the range of $-4.5$ Mm $\leq S_L \leq$ $4.5$ Mm, so as to avoid the chromospheric part of the structure.

The overall energy release profile follows a power law given in Equation (1). The larger the value of $\alpha$, the steeper the slope of the frequency distribution and hence more weight is given to the smallest heating episodes. The total energy input to the global loop remains the same for the following three simulations; namely $4 \times 10^5$ erg cm$^{-2}$ s$^{-1}$ for a total run time of $1.725 \times 10^4$ s. 

\subsection{Case A: $\alpha = 0$}
\begin{figure}
\centering
\includegraphics[width=0.35\textwidth,angle=90]{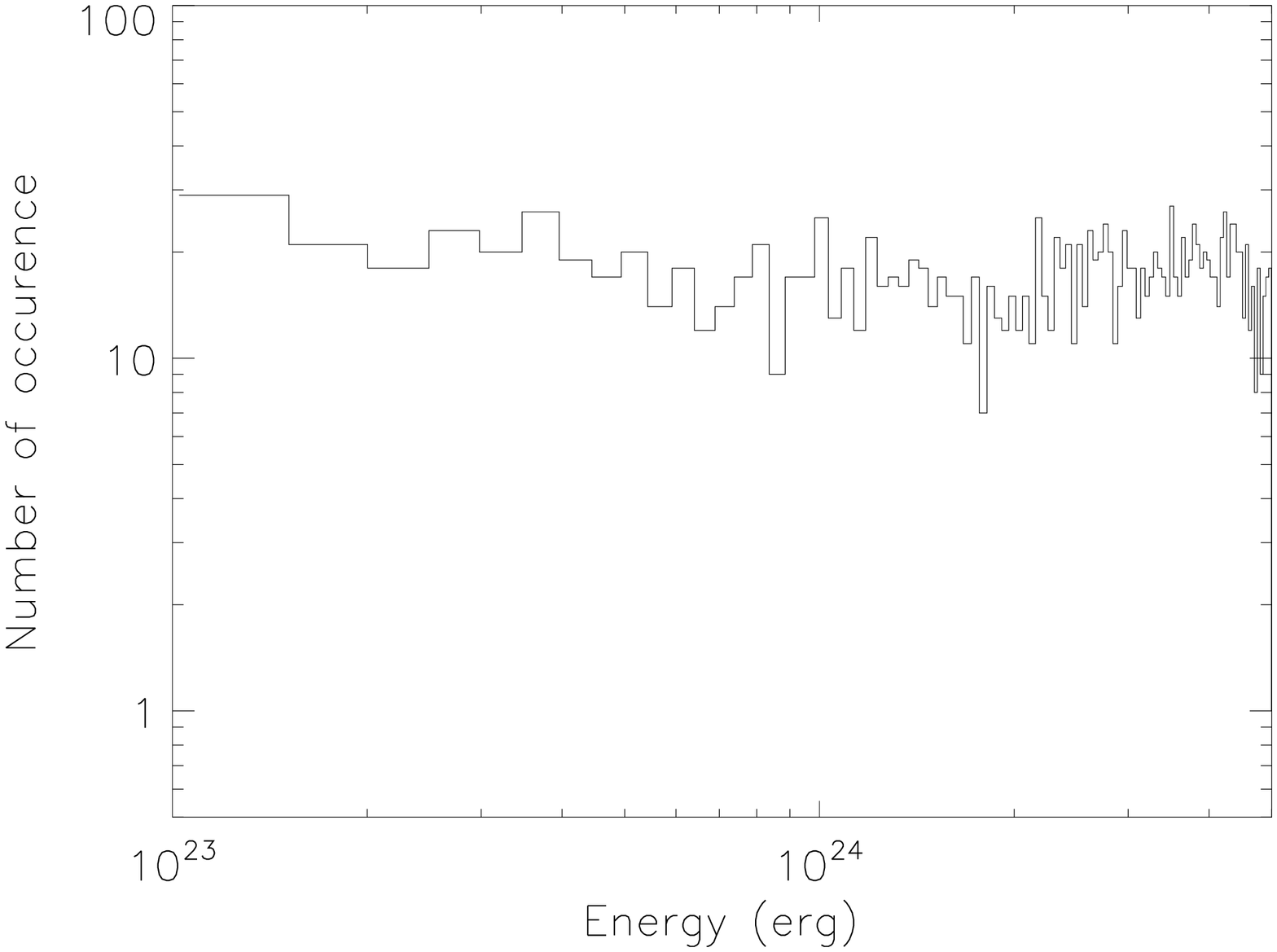}
\caption{Case A: a histogram of the size of a heating event versus the number of times it occurs during the simulation across all 125 strands; the distribution has no dominant heating scale size but is relatively uniform.\label{plotthree}}
\end{figure}

\begin{figure}
\centering
\includegraphics[width=0.55\textwidth,angle=0]{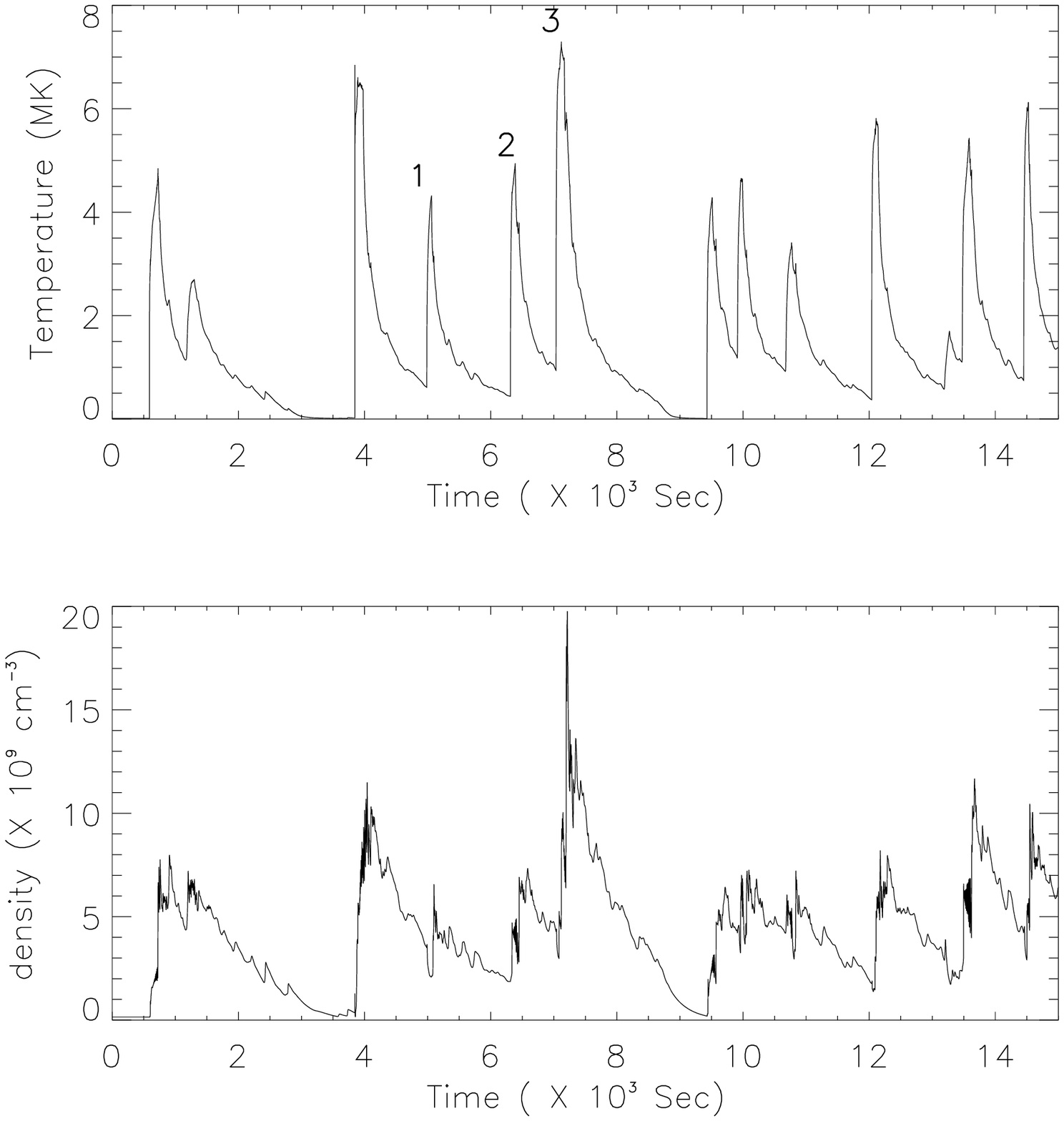}
\caption{Case A: Apex temperature (top) and density (bottom) evolution of a typical (the 55th) strand.\label{plotfour}}
\end{figure}

Firstly, consider a flat frequency distribution ($\alpha = 0$) where 1750 heating events with energy distributed randomly between $10^{23}$ and $5 \times 10^{24}$ erg provide the above said amount of total heat input. Each strand experiences on average 14 energy bursts during the simulation. 
Figure \ref{plotthree} displays a histogram of the size of an event versus the number of times it occurs during the simulation across all strands- generally, there is no preferred weighting towards any heating scale.

Let us examine the evolution of a single strand undergoing multiple heating bursts. The temperature and density evolution at the strand apex of a typical element (the 55th) is shown in Figure \ref{plotfour}; several maxima are observed in response to the reoccurring sudden energy release. As examined in Section 2, after an initial heating burst, the localised heat gets distributed by conduction and the strand plasma dissipates the excess heat through radiation unless is heated by another energy event. 

\begin{figure}
\centering
\includegraphics[width=0.65\textwidth,angle=0]{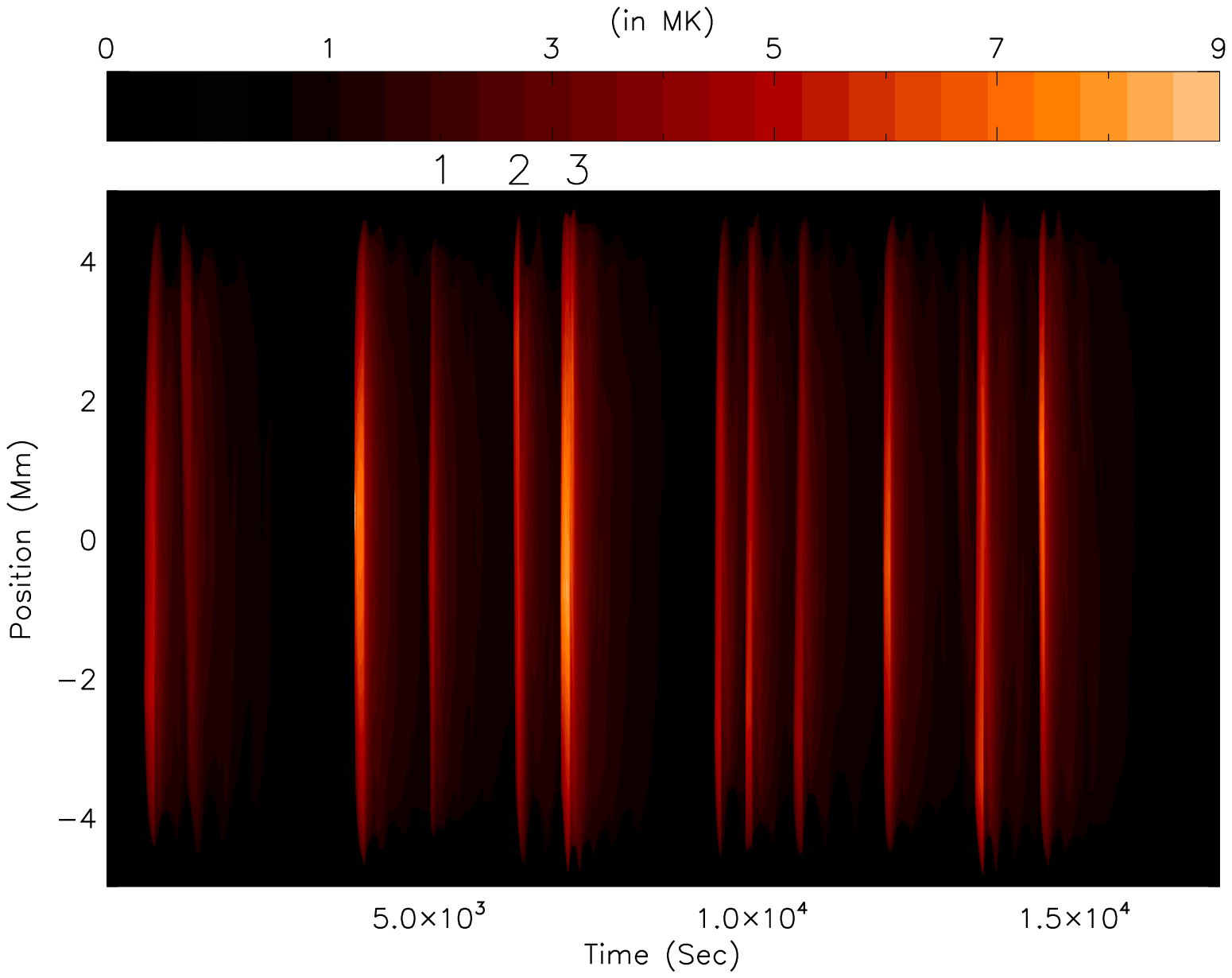}
\includegraphics[width=0.65\textwidth,angle=0]{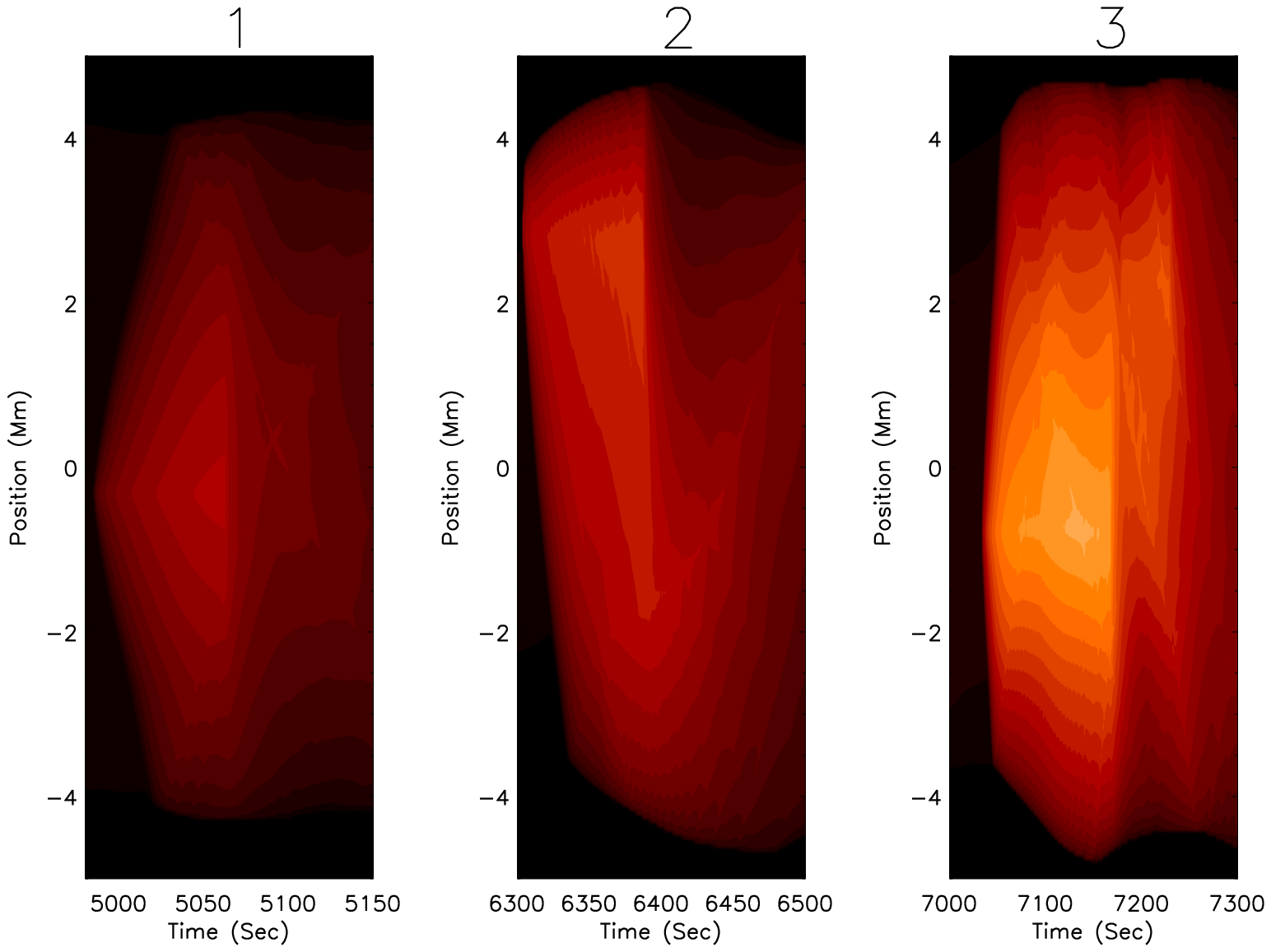}
\caption{Temperature evolution of strand 55: (top) the complete evolution for the simulation; (bottom) a close-up of 4800 to 7400 s displaying the thermal response to a number of distinct heating events. \label{plotfive}}
\end{figure}

However, what Figure \ref{plotfour} does not portray is the effect of the spatial localisation of the heating. Thus, the temperature evolution along a strand is displayed as an image plot in the Figure \ref{plotfive} (top). Each individual heating event can be identified easily as the temperature along the strand increases dramatically (sometimes up to $\sim 9$ MK); at other times, the strand plasma is much cooler. Note that, the maximum value of the temperature along the strand might not be at the strand apex.

Figure \ref{plotfive} (bottom) concentrates upon a narrower time range (4800 to 7400 seconds) and displays three specific heat deposition episodes- these are also marked in Figure \ref{plotfour}. At (1) an energy release (of $6.88 \times 10^{23}$ erg for the duration of $77.39$ s starting at $4985.25$ s) occurs at $0.05$ Mm away from the strand apex. In contrast to (1), heating event (2) occurs $\sim 3$Mm away from the apex and is larger ($2.3 \times 10^{24}$ erg released over $83.5$ s). At (3), two heating events have occurred close together in time. The first episode ($4.8 \times 10^{24}$ erg over $133.5$ s at $-1$ Mm) is released at $7032.25$ s while the second ($1.26 \times 10^{24}$ erg over $51.2$ sec, at $2.8$ Mm) is initiated at $7176$ s. The evolution of the strand temperature is clear in Figure \ref{plotfive} (bottom, right) heating the plasma to a maximum of $\sim 9$ MK. The plasma is still hot (maximum $\sim 6$ MK) when the second event occurs which hence does not have such significant impact on rising the temperature once again.

Now consider combining all 125 strands to form a ``global loop''. As individual strands are unresolved, the observed temperature has to be affected by the composite emission of all the strands together. Therefore we derive the emission measure weighted temperature ($\overline{T}_{EM}$) as

\begin{equation}
\overline{T}_{EM}=\dfrac{\sum_{i=1}^{125} n_{i}^2(s,t)dl(s)T_i(s,t)}{\sum_{i=1}^{125} n_{i}^2(s,t)dl(s)}
\end{equation}

\noindent
where $dl(s)$ is the grid resolution.

Figure \ref{plotsix} (top left) displays $\overline{T}_{EM}$ at the loop apex. Neglecting for the moment the larger drops in $\overline{T}_{EM}$ and the initial few hundred seconds as the simulation begins, the overall $\overline{T}_{EM}$ fluctuates around $1.6$ MK with an amplitude $\sim 0.4$ MK. The very low values of $\overline{T}_{EM}$ are due to how this weighted temperature is being calculated. We can see that if the density in any given strand increases dramatically, then from Equation (9), $\overline{T}_{EM}$ will be dominated by the temperature of that strand. Subsequently, Figure \ref{plotsix} (top right) displays the average apex density evolution over the 125 strands; there is a high correlation between the sharp increase in density and the sudden changes in $\overline{T}_{EM}$. Figure \ref{plotsix} (bottom) shows an image plot of $\overline{T}_{EM}$ along the loop together with an enlarged time window around $5000$s, concentrating on the propagation of a particular $\overline{T}_{EM}$ dip.

To understand better these sharp drops in the calculated $\overline{T}_{EM}$, let us concentrate upon a typical dip at around $5000$s. A reasonably sized heating event takes place in ``strand 11'' of the simulation. This strand has had enough time to relax to a cool ($\sim 10^4$K), evacuated ($\sim 10^9$ $cm^{-3}$) structure after a previous heating burst which occurred in the strand at $\sim 3000$s. It is heated again at $4904.75$s with an energy burst that lasts $113.5$s, contains $1.414 \times 10^{24}$erg of energy and is located at $3.75$Mm on the right hand side of the strand.

The detailed dynamic evolution that arises from this event can be seen in Figure 7. It is very clear that soon after the episodic heating event, a  travelling front develops rapidly from the energy release site. The reason for this drop in the $\overline{T}_{EM}$ at the loop apex is due to a low temperature ``blob'' travelling along the structure as shown in Figure 7 (bottom right).

 The local temperature rises up to 3MK with plasma being compressed ahead of the wave front. This localised cool plasma blob journeys along the strand just ahead of the corresponding increased temperature front. Subsequently, $125$s after the initial heating burst, the dense plasma front reaches the apex of the strand while at that particular instant, the local temperature is still chromospheric. Thus for this short time ($\sim 10$s) the calculated $\overline{T}_{EM}$ will produce a rapid dip. After the front passes the apex, the temperature at that location rises to coronal values. Eventually the plasma blob travels down the other strand leg, assisted by gravity.

Similar phenomenon can be observed in other strands. It appears that the development of such travelling cool plasma blobs depends on the density structure along the strand prior to the energy release as well as the location and energy content of the event itself. For example, if a heating burst is initiated when the plasma is already hot and less dense, these plasma blobs do not form. Note that these events are identical to the formation and propagation of cold plasma blobs observed in the simulation by Mendoza-Brice\~{n}o et al. (2005), who term their events microspicules.

Interestingly, these sudden dips in $\overline{T}_{EM}$ at a particular location in the loop could have observational implications. It could be the case that a loop could ``disappear'' from coronal emission and appear in transition or chromospheric ones (Schrijver, 2001; O'Shea, Banerjee, Doyle, 2007). However, it must be noted that although this phenomena consists of a flow of cold plasma along a strand, it can not be termed as classical catastrophic cooling as is analysed theoretically by, say, Karpen et al.(2001) and M\"{u}ller, Hansteen \& Peter(2003)

\subsection{Case B: $\alpha=2.29$}
After investigating Case A where the occurrence rate for all heating events has equal weighting, consider Case B where the power law index from Equation (1) is $\alpha=2.29$ (an energy event histogram similar to Figure \ref{plotthree} is displayed in Figure \ref{ploteight}a). Since we are requiring that the same total amount of energy is deposited during this simulation as in Case A, the total number of individual events occurring will increase; subsequently 7125 heating episodes take place with each strand experiencing on average 57 events.

\begin{figure}
\centering
\includegraphics[width=0.35\textwidth,angle=90]{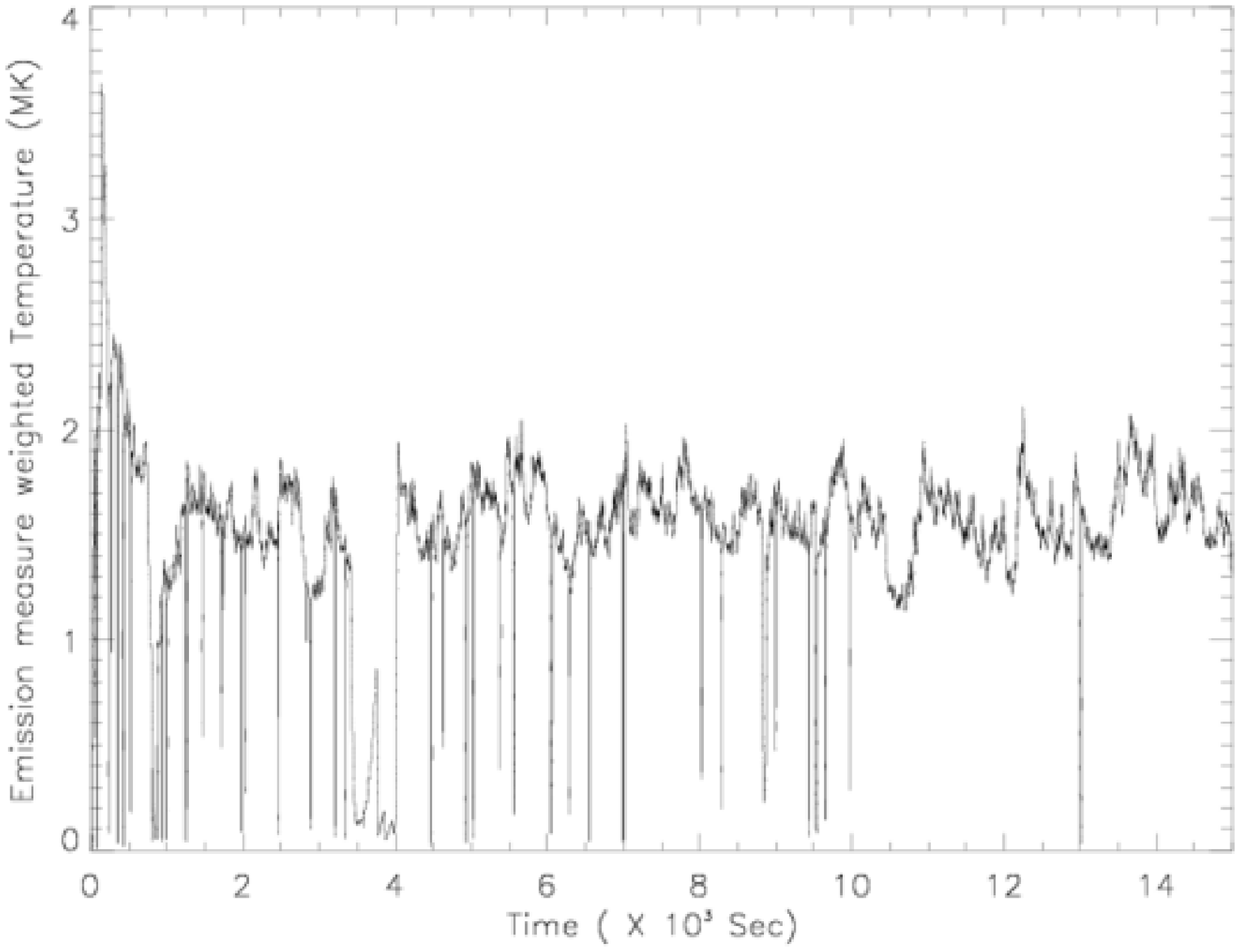}
\includegraphics[width=0.35\textwidth,angle=90]{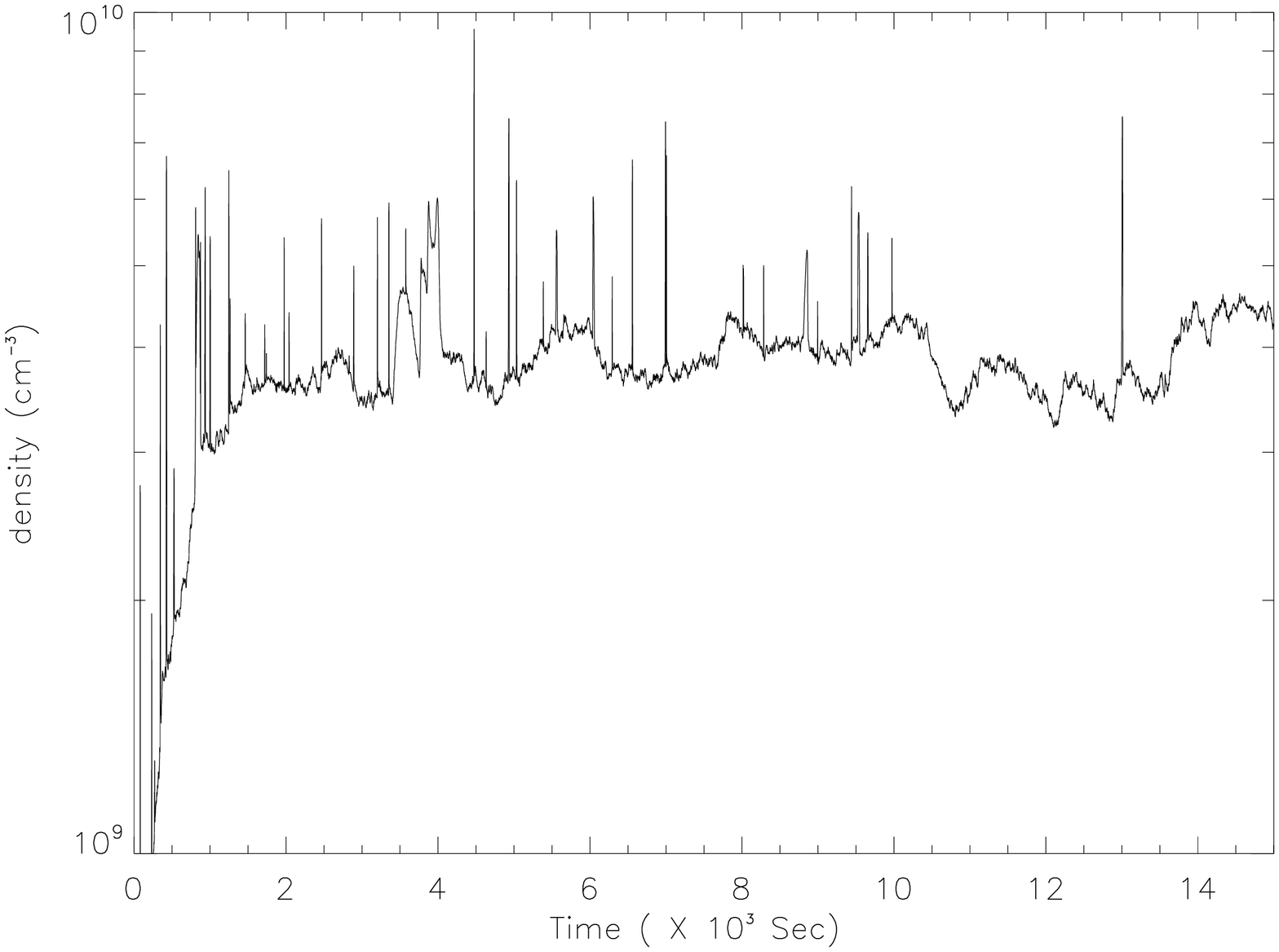}
\includegraphics[width=0.65\textwidth,angle=90]{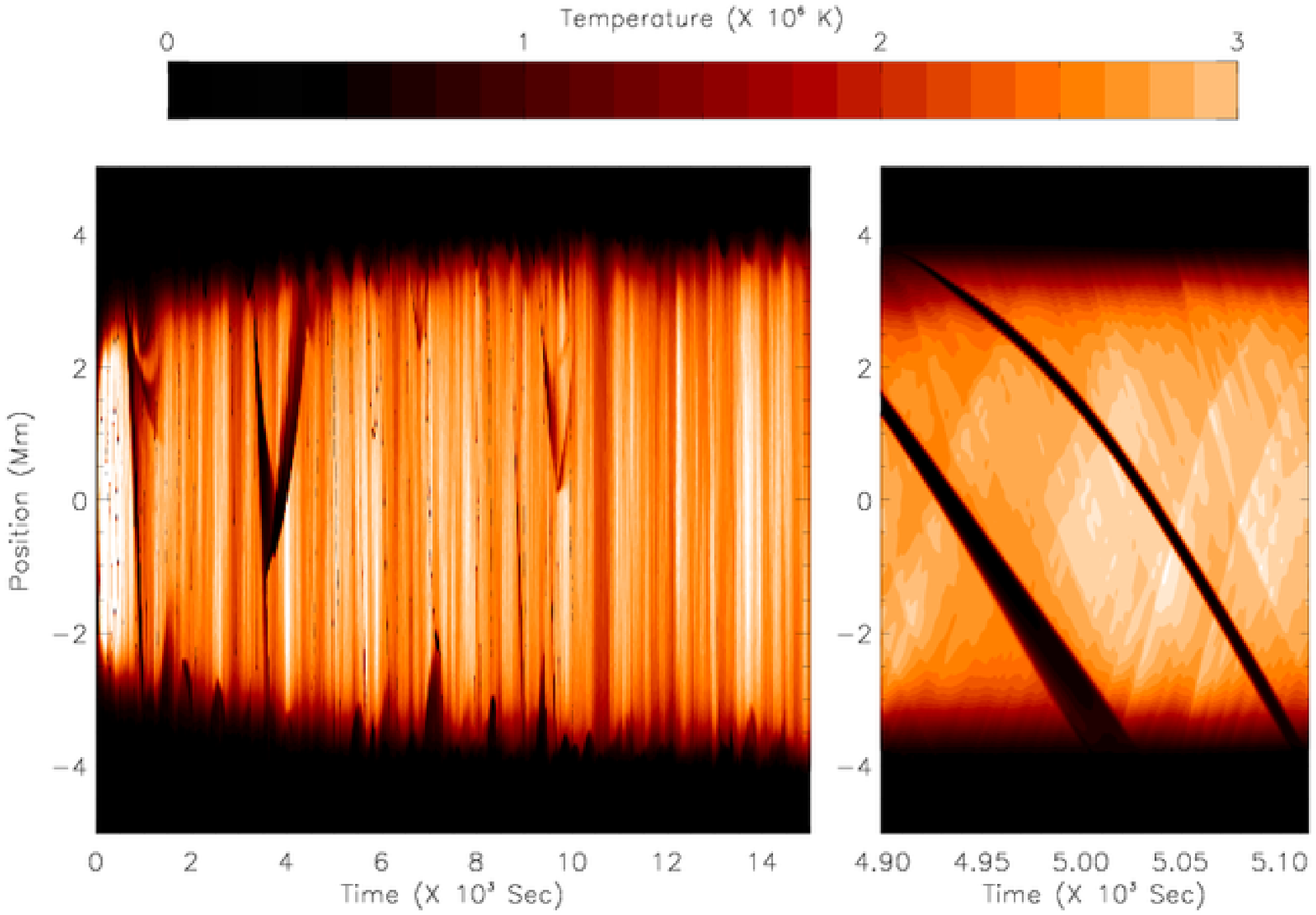}
\caption{Time evolution for 125 strands: (top left) loop apex $\overline{T}_{EM}$; (top right) corresponding loop apex density evolution; (bottom) image plot of the $\overline{T}_{EM}$ along the loop, together with a zoomed narrow time window around $5000$s.\label{plotsix}}
\end{figure}

There are a number of aspects to note. Firstly there are fewer low $\overline{T}_{EM}$ dips in Figure \ref{ploteight} (b) compared to Figure \ref{plotsix} (top left), the corresponding density evolution is also shown in Figure \ref{ploteight} (c). Secondly, the mean $\overline{T}_{EM}$ in time has increased to $\sim 2.2$ MK with a reduced fluctuation ($\sim 0.1$ MK) around this value. To explain this behaviour compared to Case A, consider once again the response of an individual strand to more numerous but less energetic bursts; this is shown for the apex temperature of a typical strand in Figure \ref{plotnine}. We can see clearly that the strand plasma is rarely provided with the opportunity to cool sufficiently before another heating burst arrives. Thus the condition for producing the cool plasma blobs is also rare (and hence the number of $\overline{T}_{EM}$ dips at the apex is also greatly reduced). For the same reason, the average strand temperature is increased throughout the simulation and subsequently, the loop temperature $\overline{T}_{EM}$ from the amalgamation of all strands also rises.

\subsection{Case C: $\alpha=3.29$}
It is instructive to investigate the effect of further increasing the value of $\alpha$. Figure \ref{plotten} (a) displays the energy event histogram for $\alpha=3.29$. Thus in this case, approximately 21500 events occur with an average of 172 heating episodes per strand.

Once again, Figure \ref{plotten}(b) plots the loop apex $\overline{T}_{EM}$; the mean value has increased slightly to $\sim 2.3$ MK compared to Case B. The bigger fluctuations in $\overline{T}_{EM}$ and density (Figure \ref{plotten}(c)) have disappeared completely. The same arguments for the absence of $\overline{T}_{EM}$ dips and for this increase in $\overline{T}_{EM}$ can be employed as outlined under Section 3.2. In particular, the strand plasma does not have adequate time to cool significantly before another heating event takes place. However, given that the energy event sizes are generally much smaller than in A and B, the impact of each event on the change in temperature is reduced. Increasing $\alpha$ further effectively ``saturates'' the temperature increase and suppresses further the fluctuations within the spatial resolution of this current simulation.

In order to obtain some idea of the ``quasi static'' thermal profile along the loop, a simple average $\overline{T}_{EM}$ profile of 125 strands is derived and the cumulative effect of combining the plasma strands over a ``typical'' 5000s period (from 5000 to 10000 s) is shown for left hand side of the loop in Figure \ref{plotten} (d). The ``usual'' overall thermal structure for a hot apex, cool footpoint loop is recovered. Figure \ref{plotten} (d) also compares this profile with two static equilibrium thermal profiles produced by the model outlined in Aschwanden and Schrijver (2002). For that particular calculation, we use a loop length of $10$ Mm and employ the apex temperature of the nano-flare heated thermal profile as a model input parameter. Note that the Aschwanden and Schrijver (2002) model has a heat input that is constant in time and that we choose to have a heating scale length that is much longer than the overall loop length, i.e., the heat input is virtually spatially uniform. In static loop profile 1, a chromosphere of length $0.4$ Mm is used- this matches the same chromosphere as employed initially in the nano-flare model. In static loop profile 2, a chromosphere of length $0.9$ Mm is chosen as at this location, the nano-flare model thermal profile begins to level off at $10^4$ K. Static loop profile 1 follows well the nano-flare thermal structure in the ``coronal part'' of the loop; however after $-2.7$ Mm, its temperature values are higher than the nano-flare model. Static profile 2 is not a good fit to the nano-flare model- from the common apex temperature, profile 2 deviates quickly from the nano-flare case, giving lower temperature values to $-3.8$ Mm and higher values in the leg. However it is to be noted that overall difference between nano-flare heated quasi static loop profile and the static loop profile 2 would be indistinguishable from current observational stand point. Similarly, although static loop profile 1 shows significant difference from the quasi static profile over $\sim 1$ Mm segment at the footpoint of the loop, considering the ambiguity of observing the loop footpoints, this difference as well could be unnoticed at the present instrumental resolution.

\begin{figure}
\centering
\includegraphics[width=0.85\textwidth,angle=90]{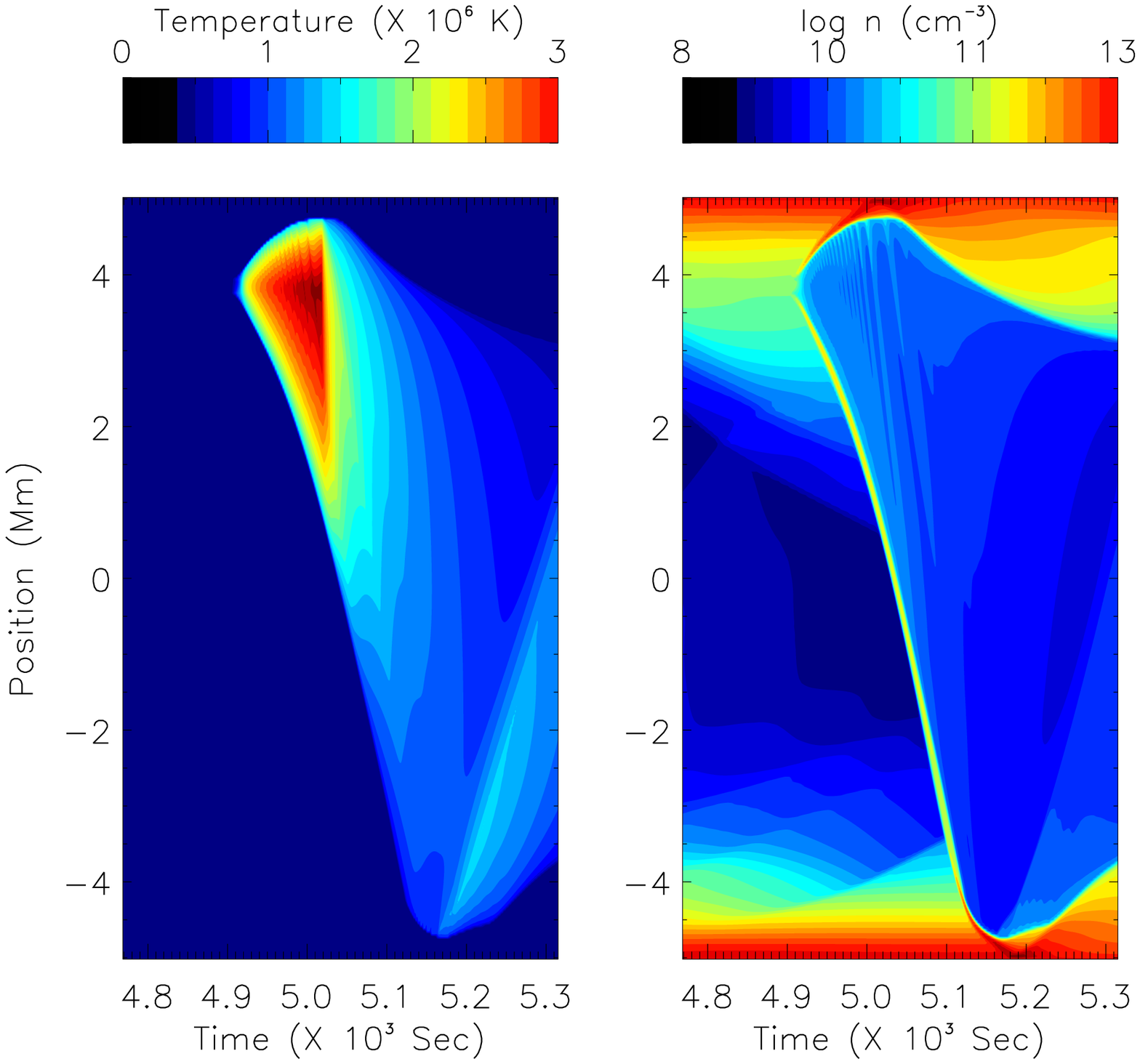}
\includegraphics[width=0.45\textwidth,angle=90]{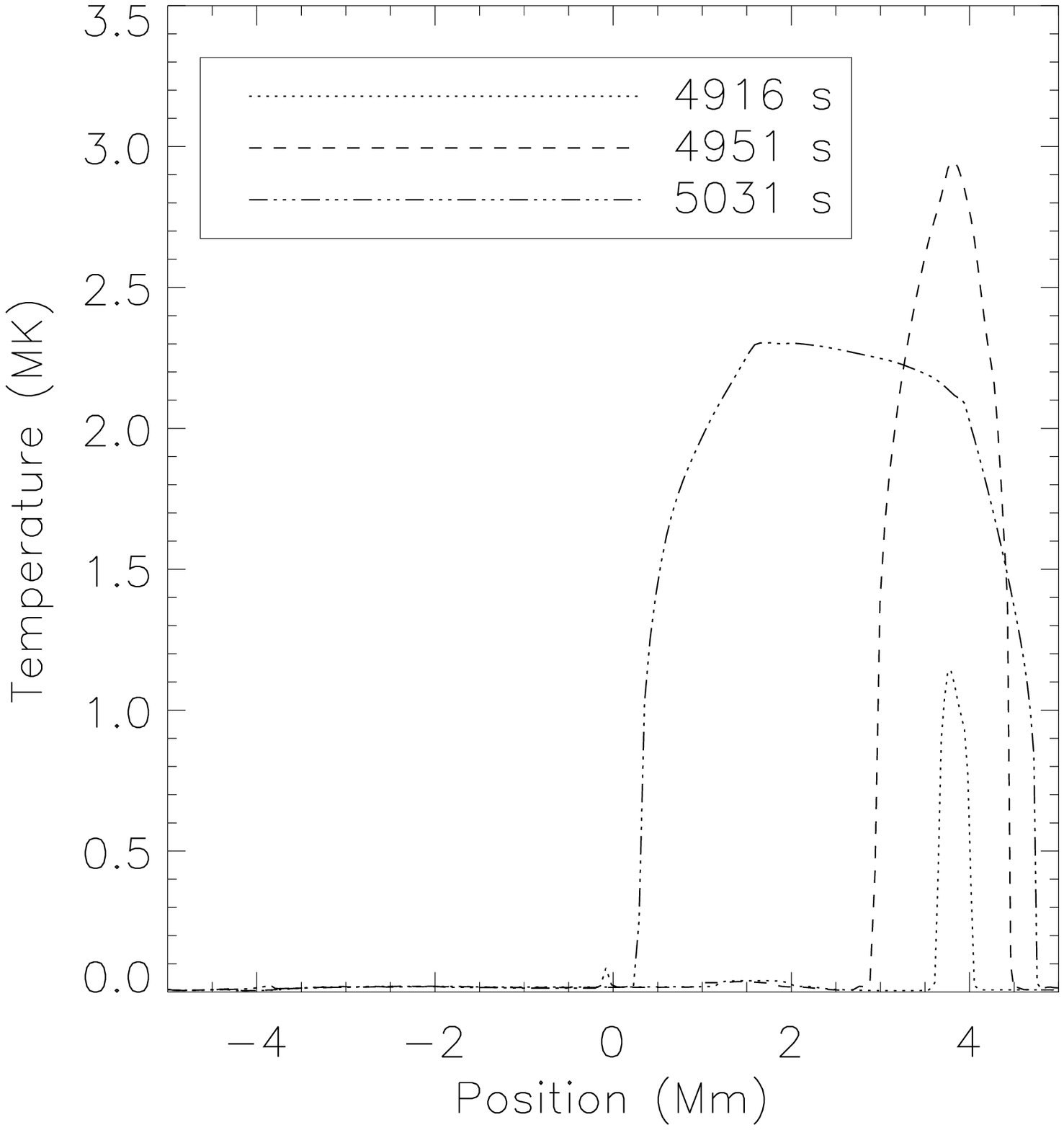}
\includegraphics[width=0.45\textwidth,angle=90]{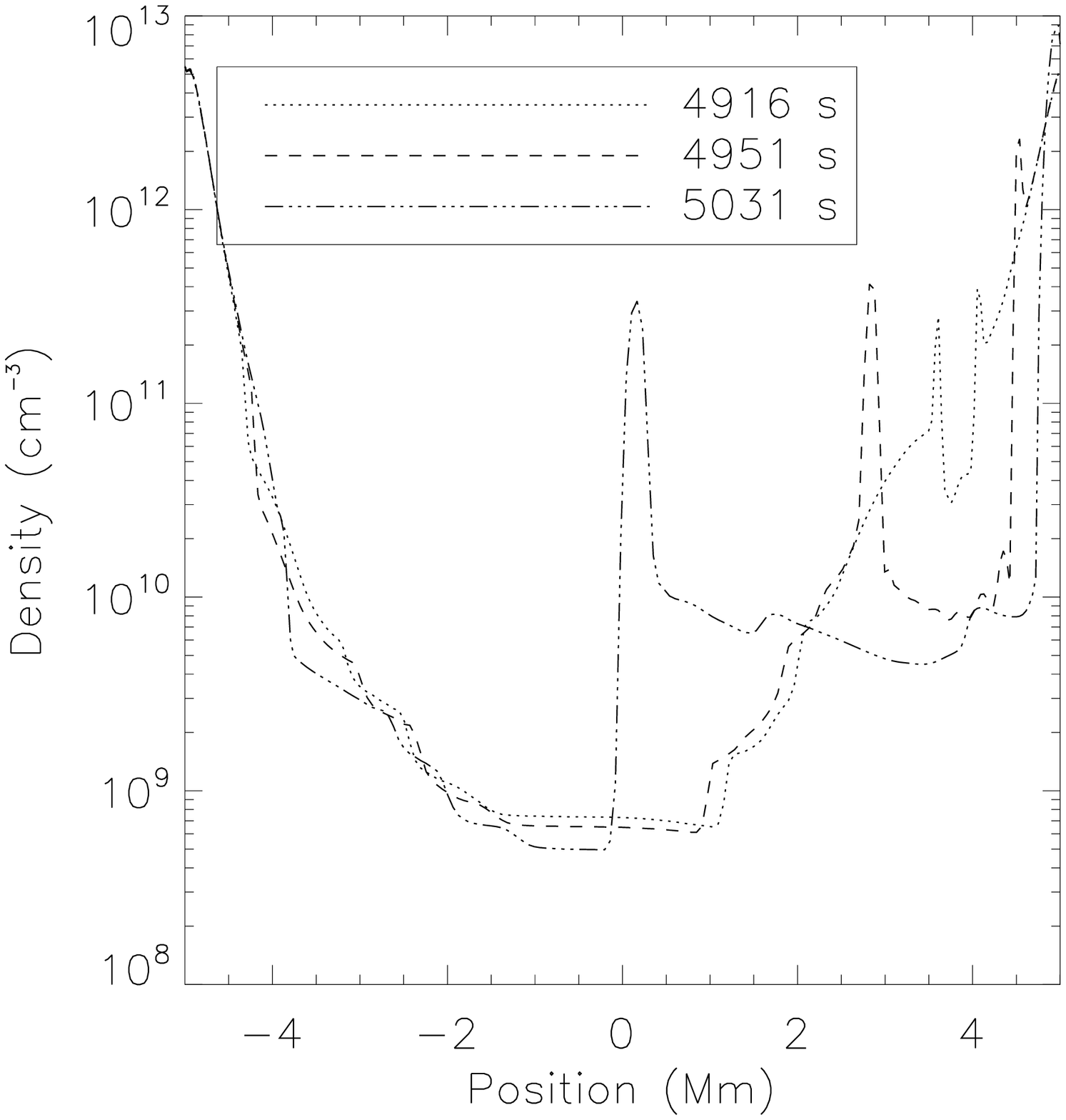}
\caption{Spatio-temporal description of the plasma blob evolution along strand 11; (top left) temperature contour before and after the dramatic event; (top right) density contour for the same time period; (bottom left) temperature profile at three different snapshots; (bottom right) density profile at the same time snapshots.\label{plotseven}}

\end{figure}

\begin{figure}
\centering
\includegraphics[width=0.35\textwidth,angle=90]{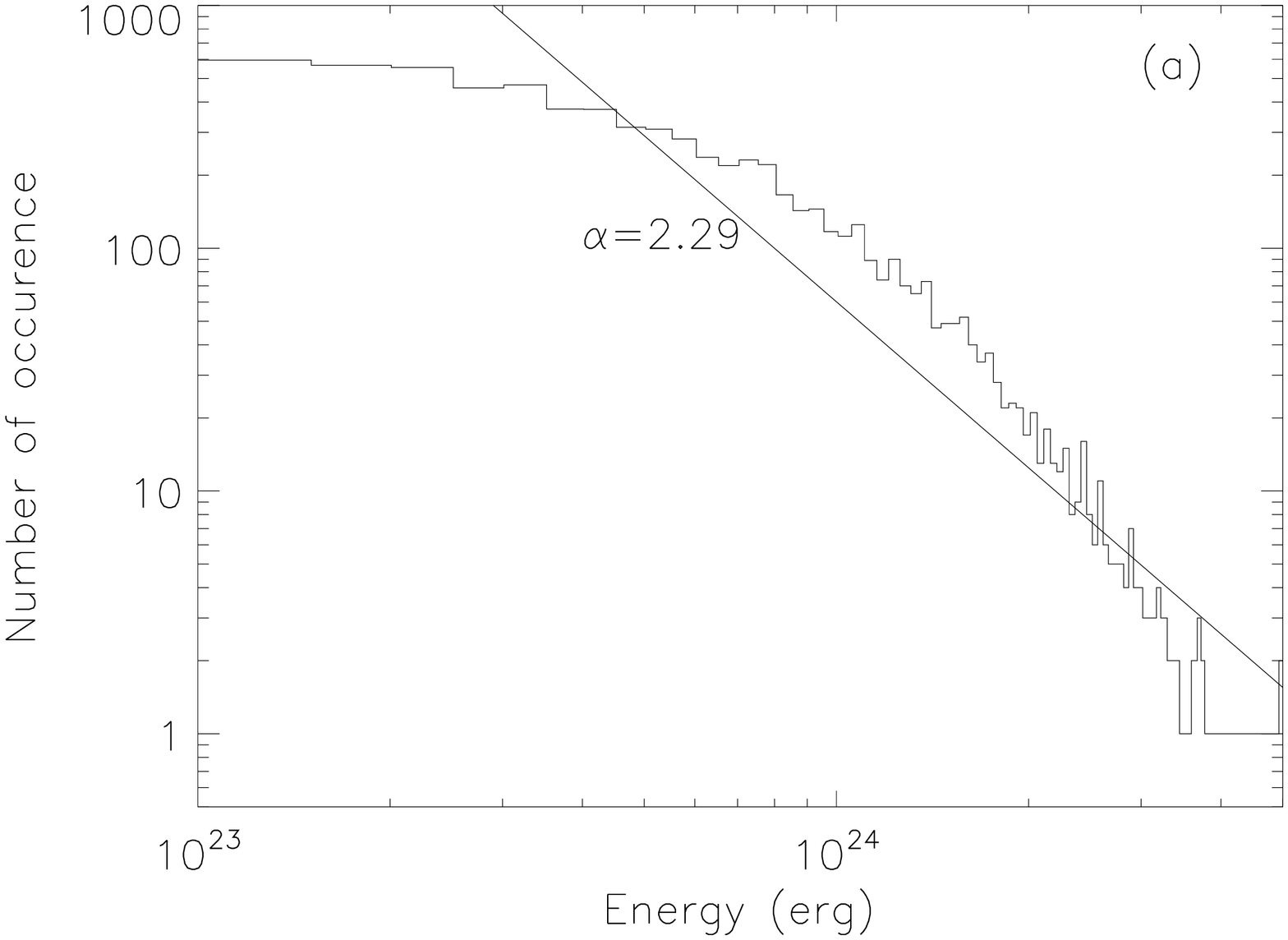}
\includegraphics[width=0.35\textwidth,angle=90]{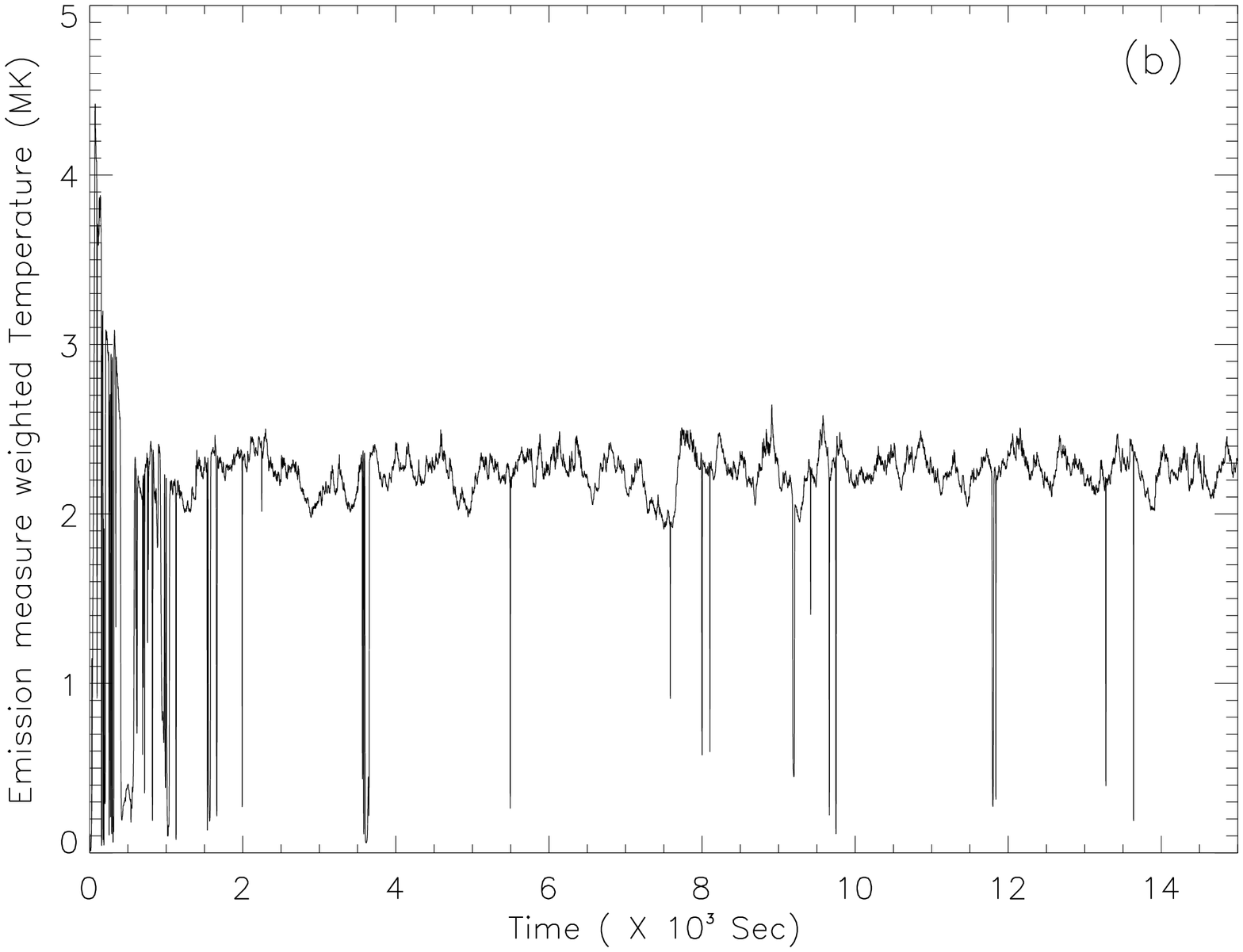}
\includegraphics[width=0.35\textwidth,angle=90]{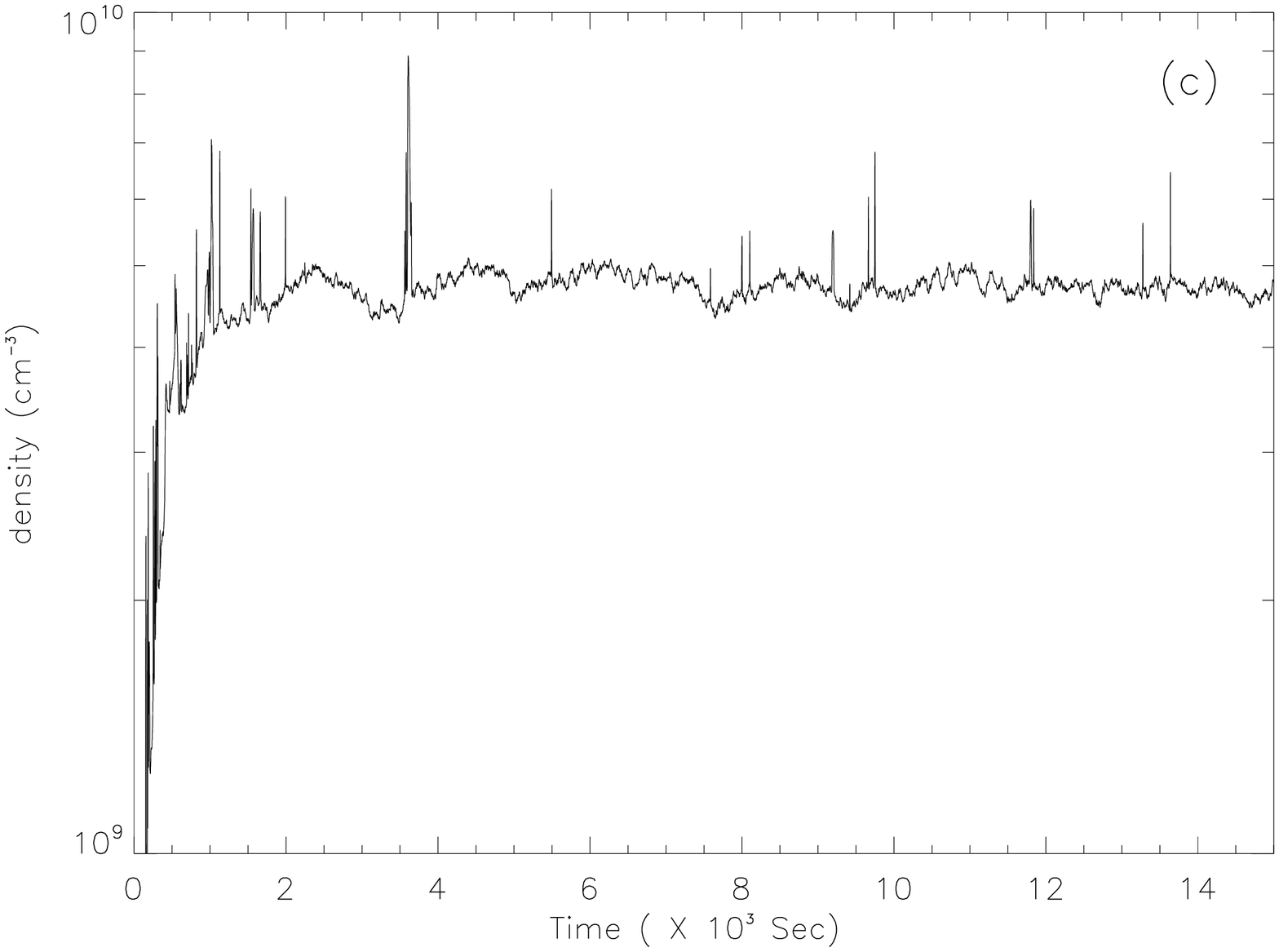}
\caption{(a) Energy histogram fitted with a straight line, to show the power law slope has a value of $\alpha=2.29$; (b) corresponding loop apex $\overline{T}_{EM}$ evolution; (c) Corresponding loop apex density evolution. \label{ploteight}}
\end{figure}

\begin{figure}
\centering
\includegraphics[width=0.35\textwidth,angle=90]{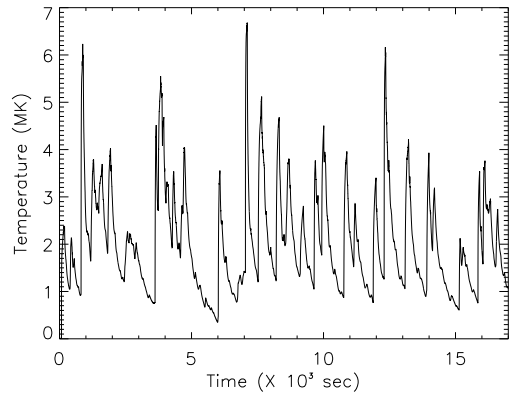}
\caption{Case B: Apex temperature evolution of a single strand when the power law slope is $-2.29$. \label{plotnine}}
\end{figure}

\begin{figure}
\centering
\includegraphics[width=0.35\textwidth,angle=90]{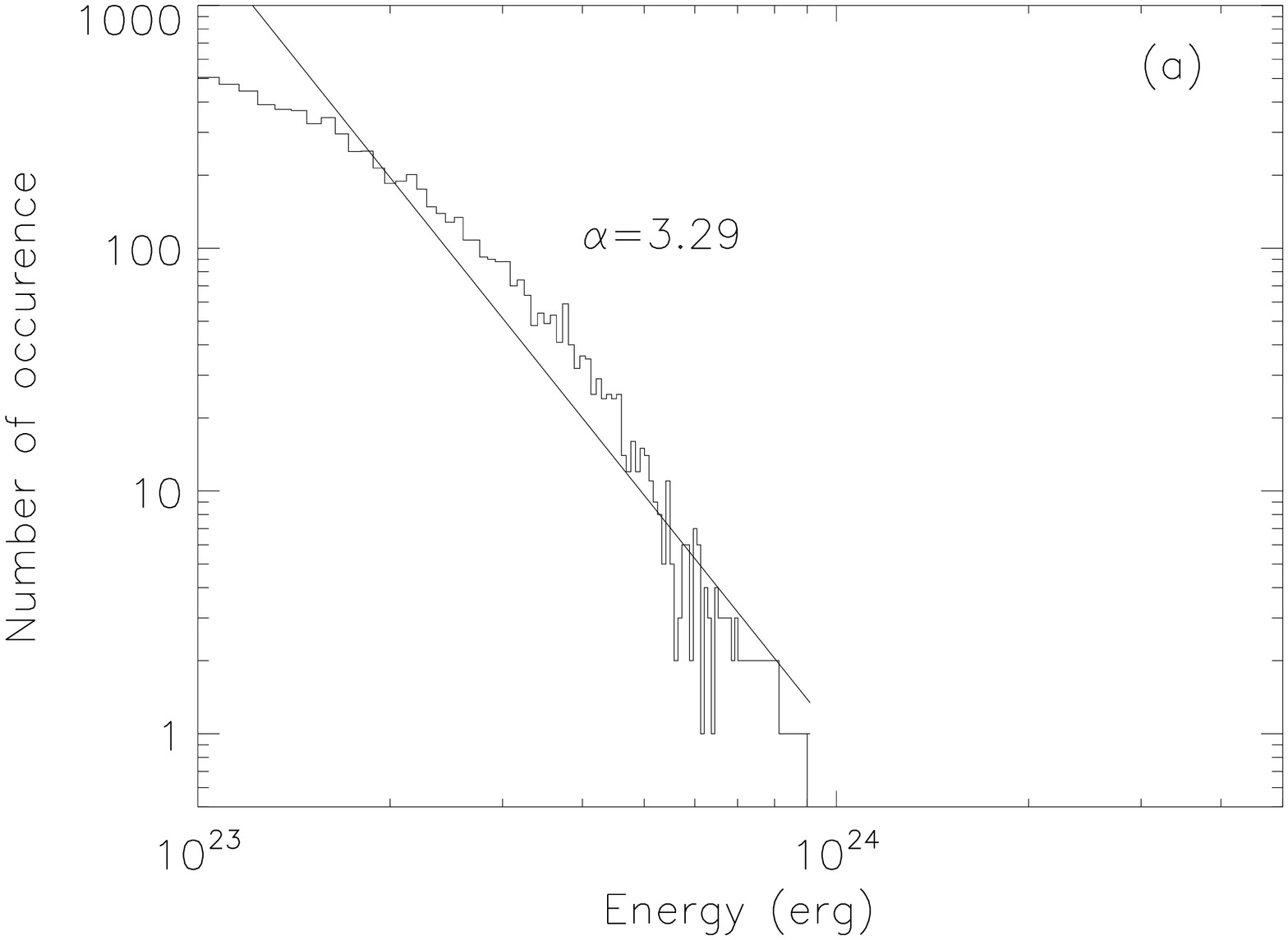}
\includegraphics[width=0.35\textwidth,angle=90]{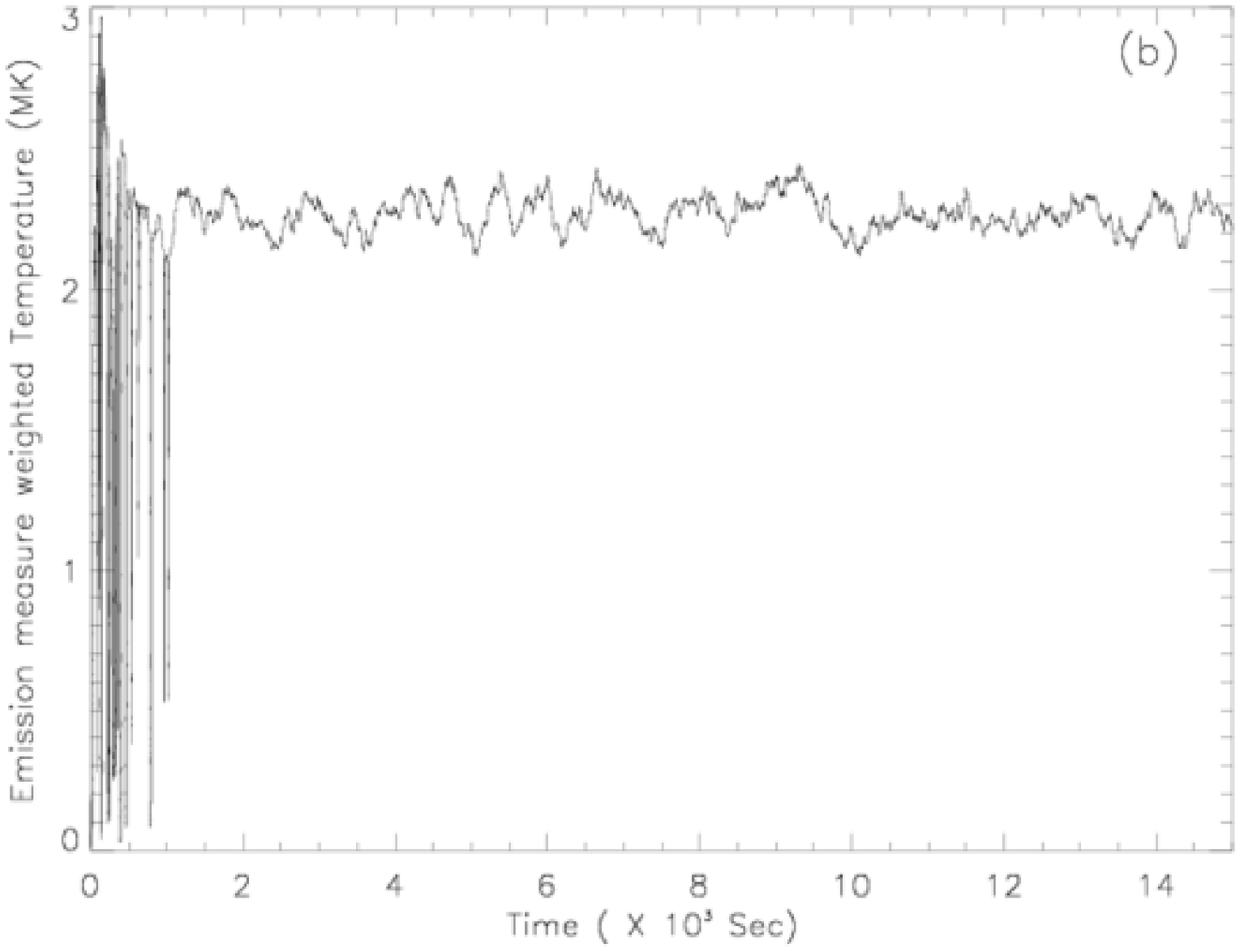}
\includegraphics[width=0.35\textwidth,angle=90]{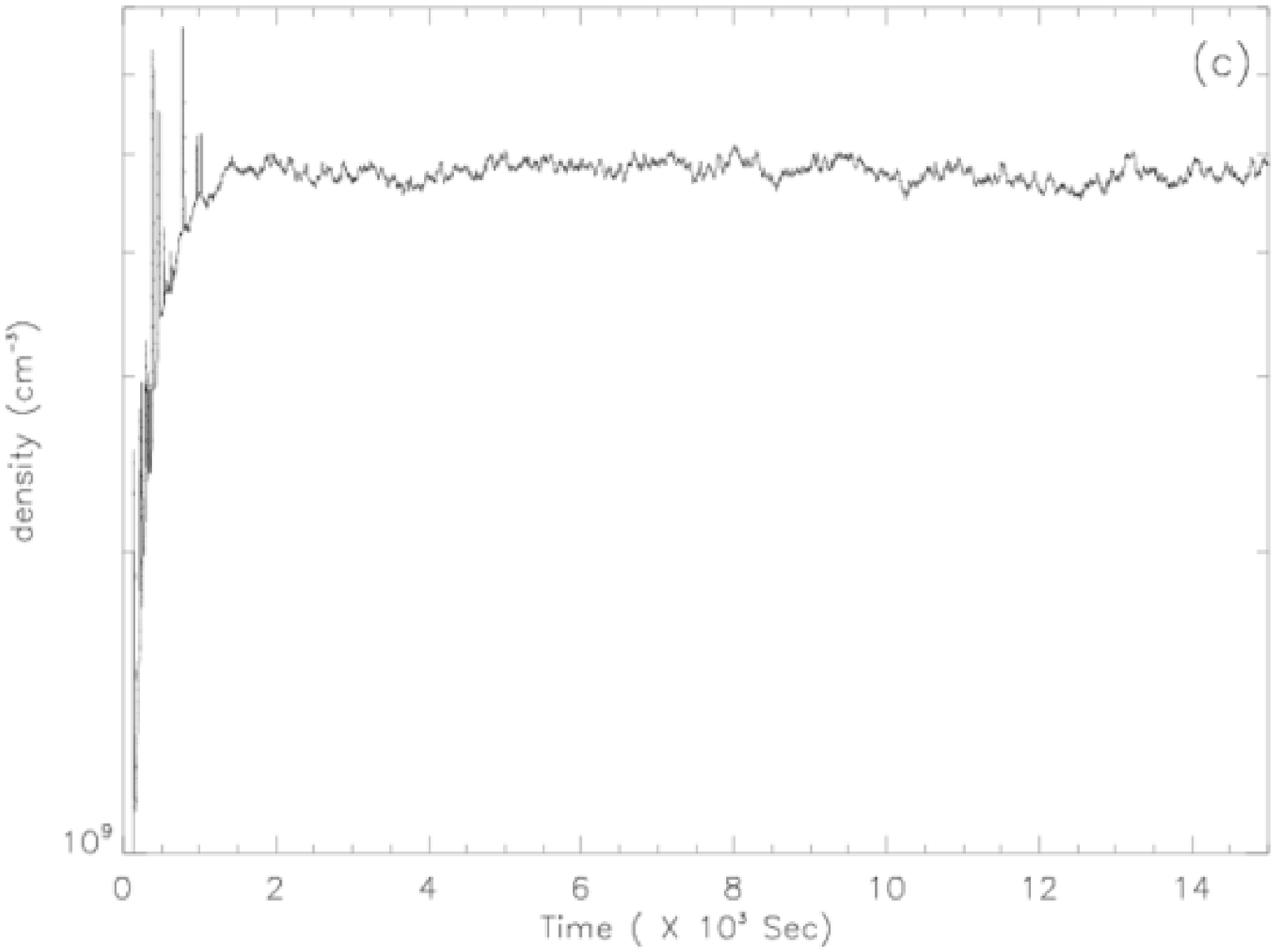}
\includegraphics[width=0.35\textwidth,angle=90]{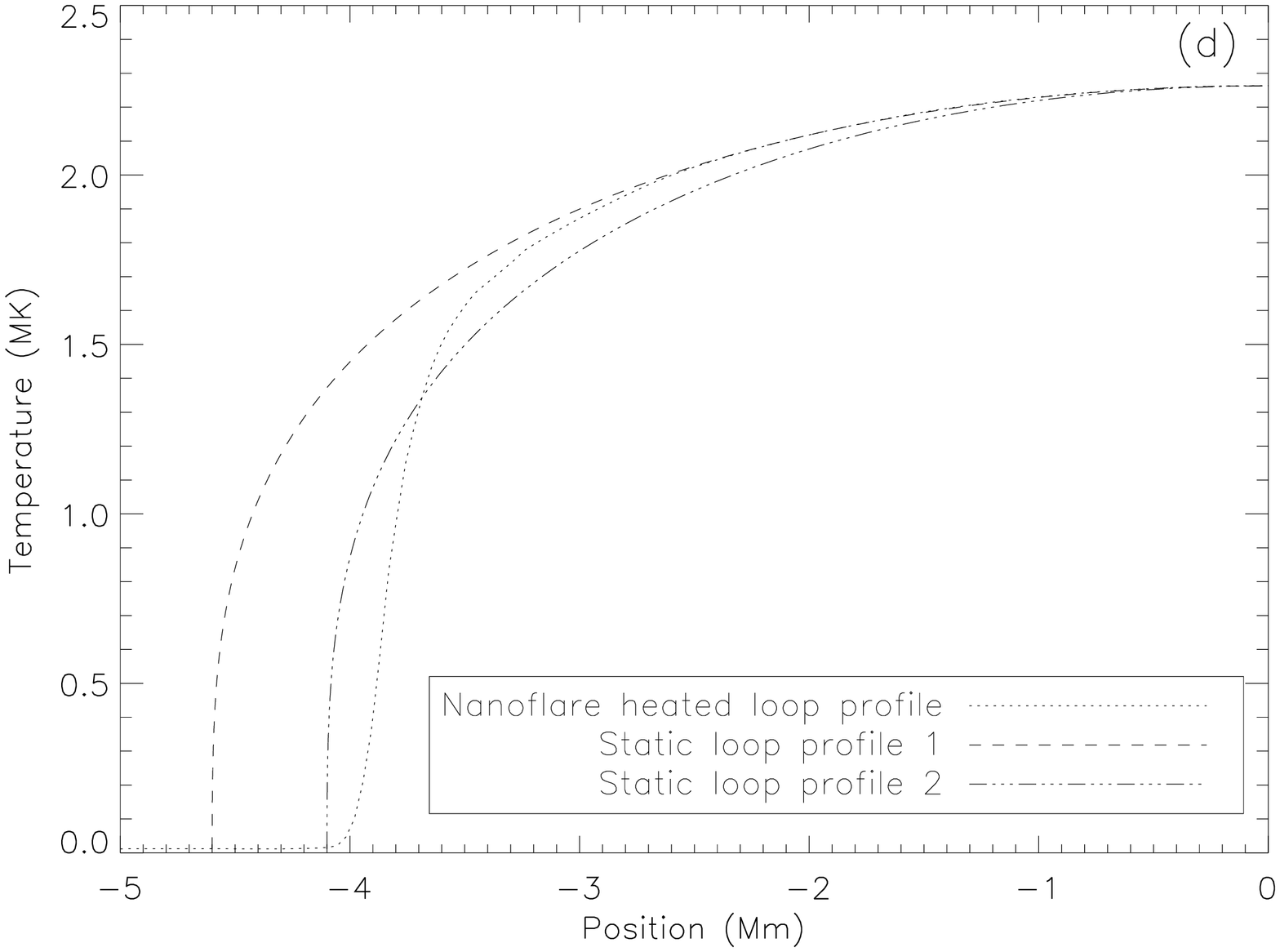}

\caption{(a) Energy histogram fitted with a straight line, to show the power law slope has a value of $\alpha=3.29$; (b) corresponding loop apex temperature evolution; (c) corresponding loop apex density evolution; (d) A comparison between nano-flare heated quasi-static half-loop temperature ($\overline{T}_{EM}$) profile and the static equilibrium temperature profile (Aschwanden, Schrijver (2002).\label{plotten}}

\end{figure}

\section{Discussion and future work}
This paper has outlined a multi-strand hydrodynamic model of a short (10 Mm) plasma loop undergoing sporadic, localised heating. During the simulations undertaken, the total energy deposited in the loop is constant while the distribution of this energy throughout the 125 individual strands follows a power law as outlined in Equation 1. The results show that even though any given strand can evolve through a series of many heating and cooling cycles, the resultant ``global loop'' thermal profile can appear relatively uniform (eg Figures \ref{plotsix}top left, \ref{ploteight}b and \ref{plotten}b). Increasing the value of the power law index ($\alpha$) in Equation 1 (from 0 to 2.29 to 3.29), increases the subsequent mean loop apex temperature though, at the current simulation resolution, this temperature value saturates if $\alpha$ is increased further.

Note that for computational expediency, these investigations have considered a short 10 Mm loop with the limited filamentation of $N_{s}=125$ strands. If the number of strands was increased further ($N_{s}>125$), the total volume occupied per strand would decrease. Concentrating upon Case B ($\alpha=2.29$), if the total energy deposition into the loop throughout the simulation remains fixed and the lower energy range cut-off ($E_{low}$) does not change (from Figure \ref{plotthree} that is set currently at $E_{low}=10^{23}$ erg), then the number of events per strand will decrease but the ``energy density per event per strand'' will increase. That is, although there would be less events occurring per strand throughout the simulation time, the impact on the temperature evolution will be greater due to the reduced individual strand volume. Cargill (1994) found that for the 0-D model, increasing $N_{s}$ leads to a slight overall increase in average loop temperature. Further investigations are underway to quantify fully the effect of altering $N_{s}$ in this hydrodynamic model.

Also, if the overall global loop is lengthened, then subsequently longer strands will have longer conductive cooling times. However, the total energy deposited within the complete simulation time frame would need to increase accordingly (approximately tenfold for a $100$Mm loop) to be able to raise the temperature of the longer strands to values comparable to this current simulation. Considering Case B  once again, this could be achieved with this specific power law index by either (i) keeping the original energy range (of $10^{23}$ to $5 \times 10^{24}$ erg) fixed over which the events can occur but simply allowing an increased number of each event size to take place; or (ii) widening the event range so that many, much smaller events ($< 10^{23}$ erg) and a few much larger events ($> 5 \times 10^{24}$ erg) can happen. It is difficult to estimate fully the impact of each of these different ``energy scenarios'' on the overall plasma evolution. For the former, it could be envisaged that the resulting thermal evolution, average apex temperature and apex temperature variation would be quite similar to that already outlined in Case B. For the latter, the extension to much smaller, more numerous energy events, would mean that the plasma reacts similar to Case C (a slightly raised apex temperature compared to (i) and a reduced apex temperature variation). Once again, simulations are being undertaken to examine the consequences on longer loop structures. 

Also it would be useful to forward fold the plasma parameters through the instrument response functions of say, TRACE and Hinode XRT. In particular, the high temperature of individual strands that should be observable in the XRT lines could lead to important distinguishing factor for coronal heating diagnostics; in contrast, TRACE EUV emission could come mainly from plasma that is cooling into the passbands. This is where this multi-strand modelling approach has a distinct advantage over the 0D models as the current simulations can be compared directly to the observed dynamics/ signatures along individual loop structures. These aspects will be tackled thoroughly in a future paper.

\section{Acknowledgments}

This work was supported by a Science and Technology Facilities Council
Standard Grant (PP/C502506/1). The authors wish to thank T. Arber for
providing the original well-documented Lagrange-remap numerical code and the referee for helpful and constructive suggestions on improving the manuscript.

\end{document}